\documentclass[useAMS,usenatbib]{mnras}
\usepackage{amsmath}
\usepackage{mathtools}
\usepackage{amssymb}
\usepackage{graphicx}
\usepackage{multirow}
\usepackage[dvipsnames]{xcolor}
\usepackage{array}
\usepackage{subcaption}
\usepackage[export]{adjustbox}
\usepackage{tabularx} 
\usepackage{comment}
\usepackage[para,online,flushleft]{threeparttable}

\title[Hybrid corona and soft lags in Fairall~9]{Hybrid corona and transient soft X-ray lags in Fairall~9}

\author[K. Khanthasombat et al.]{K. Khanthasombat$^{1}$, P. Chainakun$^{1,2}$\thanks{E-mail: \href{mailto:pchainakun@g.sut.ac.th}{pchainakun@g.sut.ac.th}}, W. Luangtip$^{3,4}$, J. Jiang$^5$, A. J. Young$^6$ \\
$^1$School of Physics, Institute of Science, Suranaree University of Technology, Nakhon Ratchasima 30000, Thailand\\
$^2$Centre of Excellence in High Energy Physics and Astrophysics, Suranaree University of Technology, Nakhon Ratchasima 30000, Thailand\\
$^3$Department of Physics, Faculty of Science, Srinakharinwirot University, Bangkok 10110, Thailand\\
$^4$National Astronomical Research Institute of Thailand, Chiang Mai 50180, Thailand \\
$^5$Department of Physics, University of Warwick, Gibbet Hill Road, Coventry CV4 7AL, UK\\
$^6$H.H. Wills Physics Laboratory, University of Bristol, Tyndall Avenue, Bristol BS8 1TL, UK}
\date{Accepted XXX. Received YYY; in original form ZZZ}

\pubyear{2025}

\begin{document}
\label{firstpage}
\pagerange{\pageref{firstpage}--\pageref{lastpage}}
\maketitle

\begin{abstract}

Fairall~9 is among the most massive Seyfert galaxies exhibiting a strong soft X-ray excess, but it is challenging to probe soft X-ray reverberation lags (if any) due to the long intrinsic timescales expected from its large black hole mass of $\sim 2.55 \times 10^8 M_\odot$. We fit five \emph{XMM-Newton} spectra of Fairall~9 using the hybrid {\sc reXcor} model taking into account both hot and warm corona. The soft excess is explained by a combination of a physically motivated warm corona and the disc reflection. Then, we perform a wavelet coherence analysis of the light curves between 0.3--1 and 1--4~keV bands. The spectral fits are consistent with a rapidly spinning black hole ($a = 0.99$), a warm corona with optical depth $\sim$10--30, and a hot lamp-post corona located at either $5$ or $20~r_{\rm g}$. This configuration supports a coexisting hot and warm corona scenario, allowing the disc to extend almost to the event horizon. Our wavelet analysis on combined observations reveals signatures of transient soft X-ray lags, confined to specific time-frequency intervals. The earlier observations exhibit more variable and transient lag behavior. In contrast, the later observations display more persistent soft X-ray lags at the frequencies of $\sim 9\times 10^{-6}$--$2.5 \times 10^{-5}$~Hz, with amplitudes reaching $\sim$1000~s. The results indicate a progressively stable disc-corona configuration in later observations. Given the mass and geometry of Fairall~9, the observed soft lags appears plausibly consistent in both size and timescales with expectations from X-ray reverberation.

\end{abstract}


\section{Introduction}

Active Galactic Nuclei (AGNs) are among the most luminous objects in the Universe, powered by accretion of matter onto supermassive black holes residing at galactic centers. Their radiation spans the full electromagnetic spectrum, with X-ray emission often dominating the observed flux. This X-ray emission is attributed to inverse Compton scattering of thermal ultraviolet (UV) and optical photons from the accretion disc by a population of relativistic electrons in a compact, hot corona situated in the immediate vicinity of the SMBH \citep{Shakura1973, George1991, Fabian2000}. The resulting power-law continuum with the energy cut-off $\sim$ 100 keV\footnote{This can be characterized by the coronal temperature of $\sim$ 50 keV depending on the properties (e.g., optical depth) of the corona.} is a hallmark of AGN X-ray spectra. A portion of X-rays are reprocessed by the accretion disc, producing the reflection spectrum, including key features such as the Fe-K$\alpha$ fluorescence line at $\sim 6.4$ keV and the Compton reflection hump \citep[e.g.,][]{Reynolds2003, Garcia2014}.

Fairall~9 \citep{Fairall1977} is a nearby Type 1 AGN at redshift $z = 0.047$ \citep{Emmanoulopoulos2011}, hosting a massive central black hole of $\sim (2.55 \pm 0.56) \times 10^8 M_\odot$ \citep{Peterson2004}. 
\emph{Swift} monitoring has shown that X-ray and UV/optical variations are strongly correlated, with UV/optical lagging behind X-rays by $\sim 2$--10~days, suggesting that the majority of the optical/UV emission arises from X-ray reprocessing \citep{Pal2017}. The behavior of longer wavelengths lagging behind shorter ones was also reported by \citet{Buisson2017}. However, a comprehensive data-driven reverberation model connecting UV/optical and X-ray variability is still lacking \citep{Hagen2023-Feb}. Furthermore, a recent study by \citet{Edelson2024}, based on a different \emph{Swift} campaign covering 1.8 years of near-daily monitoring, found that X-ray flux variations correlate poorly with UV/optical variations. When the light curve was divided into four $\sim$60-day segments, the UV/optical lag spectrum remained highly stable across all periods. In contrast, the X-ray to UV lags displayed substantial instability, with both the magnitude and sign of the lags changing between segments. The work of \cite{Partington2024}, using \emph{NICER} and \emph{Swift} observations, showed that UV variations on longer timescales ($>30$ days) are possibly intrinsic to the accretion disc, whereas shorter-term UV variations can be attributed to X-ray reprocessing, and the generally weak X-ray/UV correlation suggests a dynamically evolving corona. This suggests that the X-ray to UV relationship is more complex than the simple reprocessing model commonly used for AGN.  

Nevertheless, Fairall~9 is unobscured in X-rays, suggesting the absence of a warm absorber along the line of sight \citep{Emmanoulopoulos2011}. These properties make Fairall~9 particularly well-suited for studies the disc-corona interaction. This source also shows strong soft X-ray excess below $\sim$2 keV, which is a longstanding feature seen in many type 1 AGN \citep[e.g.,][]{Crummy2006, Gliozzi2020}. In addition to a power-law continuum with cold and ionized reflection, the soft excess may require an additional Comptonization component, i.e., a spatially distinct Comptonizing zone \citep{Lohfink2012, Lohfink2016}. 

Many recent studies suggest that the soft excess can be explained by a warm, optically thick Comptonizing region, known as a warm corona, that upscatters optical/UV disc photons into the soft X-ray band \citep[e.g.][]{Kubota2018, Petrucci2018, Porquet2018, Ursini2020, Xu2021, Xiang2022, Ballantyne2024}. This warm corona model has gained strong observational support, offering a physically plausible mechanism that avoids the extreme conditions often required by reflection-only scenarios. 

It is well established that, within the X-ray reflection framework, X-ray reverberation lags on inner-disc reflection timescales can arise from the light travel time between the corona and the reflecting regions of the accretion disc \citep{Uttley2014,Cackett2021}. However, using \emph{Suzaku} data, \citet{Yaqoob2016} found that the Fe-K emission in Fairall~9 appears narrow and originates from regions thousands of gravitational radii away, suggesting either a low-spin black hole or a truncated or non-illuminated inner disc. In contrast, combined \emph{XMM-Newton} and \emph{NuSTAR} observations indicate that the inner edge of the accretion disk lies very close to the black hole, implying an extremely high black hole spin \citep{Lohfink2016}. To date, no soft X-ray or Fe-K reverberation lags have been reported on timescales associated with inner disc reflection. This absence aligns with expectations, as the large central mass in Fairall~9 results in long X-ray reverberation timescales (tens to hundreds of kiloseconds), that fall below the frequency range most effectively probed by the available X-ray data, such as those in the \emph{XMM-Newton} archive. Furthermore, the weak inner-disc reflection diminishes the amplitude of potential reverberation signals, making them more difficult to detect.

In this work, we apply the lamp-post geometry with a {\sc reXcor} warm corona model \citep{Xiang2022, Ballantyne2024} to fit the X-ray spectrum of Fairall~9. Our goal is to constrain the physical parameters of the accretion geometry by employing a hybrid model that incorporates both warm Comptonization and relativistic reflection, allowing us to disentangle their respective contributions to the soft excess and the interpretation of X-ray lags. 

To further investigate time variability in Fairall~9, we perform wavelet analysis by examining wavelet coherence and phase differences between light curves in two energy bands: soft 0.3--1 keV and hard 1--4 keV. The soft band seems to be influenced by reflection processes, with a possible warm-Comptonization contribution, while the hard band is generally continuum-dominated. We examine wavelet lags for individual observations as well as for combined, stitched light curves, and perform simulations to illustrate wavelet-lag behavior and potential artificial features within the X-ray reverberation framework. We aim to determine the signature of soft X-ray lags, i.e. the variations in reflection-dominated soft band lagging behind the continuum-dominated hard band due to the different light travel distances. These lags may themselves be variable. Unlike Fourier-based methods previously used to model and analyze X-ray reverberation lags \citep[e.g.][]{Fabian2009, Wilkins2013, Cackett2014, Emmanoulopoulos2014, Chainakun2016, Kara2016}, wavelet techniques are well-suited for detecting transient or frequency-dependent phase lags that may not persist throughout the entire observation \textcolor{red}{\citep{Ghosh2023, Wilkins2023}}. By retaining both temporal and frequency information simultaneously, this approach enables the detection of non-stationary features that are otherwise averaged out in Fourier space. This is particularly useful in AGN studies, where the variability timescales and physical conditions can change over the course of a single observation. In such cases, soft lags may originate from localized, non-stationary reverberation events or from stationary reverberation signals that become partially obscured during certain intervals due to fluctuations in data quality.  


The wavelet-lag analysis of AGN has been investigated before by \cite{Wilkins2023} in the well-known narrow-line Seyfert 1 galaxy IRAS~13224--3809. The main focus of our work is to apply wavelet methods to Fairall~9. By extending the methodology to this AGN, we provide complementary evidence that broadens the applicability of wavelet-based reverberation studies. Note that IRAS~13224--3809 hosts a significantly smaller black hole, estimated at $\sim 2 \times 10^6M_\odot$ \citep{Alston2020}. This lower mass leads to soft X-ray reverberation lags occurring at much higher frequencies \citep[e.g.,][]{Kara2013}. The X-ray lags in IRAS~13224--3809 also thoroughly studied using traditional lag-frequency techniques \citep{Alston2020, Caballero2020, Hancock2023}, power spectral density (PSD) analysis \citep{Chainakun2022a, Mankatwit2023}, and Granger causality tests \citep{Chainakun2023, Nakhonthong2024}. Using both simulated data and XMM-Newton observations, we explore whether combining light curve stitching with wavelet coherence and lag analysis can extend the temporal frequency range accessible for AGN variability studies, particularly in cases where individual observations are relatively short, such as for Fairall~9. We then discuss the wavelet lag results of Fairall~9 alongside those of IRAS 13224--3809. Finally, we examine the mass-scaling of reverberation signatures and assess whether the wavelet-derived lags are consistent with expectations based on disc-corona light travel time delays.

The rest of this paper is structured as follows. Section~2 provides details of the observational data, including the procedures for extracting X-ray spectra and light curves. In Section~3, we review the warm corona spectral model and describe the key physical parameters that define the model components. The wavelet-lag analysis is presented in Section~4, followed by simulation tests in Section~5. Then, the main results are presented in Section~6, including spectral fitting outcomes, parameter constraints, and the variability properties derived from our analysis of Fairall~9. We discuss the implications of these findings in Section~7, and conclude the results in Section~8.

\section{Observations and data reduction}

All Fairall~9 data used in this work were observed by {\it XMM-Newton} \citep{Jansen2001} in which its data were obtained from {\it XMM-Newton} Science Archive\footnote{\url{https://nxsa.esac.esa.int/}}; the observational detail are listed in Table~\ref{tab:xmm_obs}. To obtained the source's science products, the observation data files (ODF) were reprocessed using the software {\it XMM-Newton} Science Analysis System ({\sc xmm-sas}) version 21.0.0 running with the lasted calibration files at the time that the data were reprocessed. Here, we used only the pn data to obtain the best data quality. In brief, the task {\sc epproc} with its default parameters was excused to create the reprocessed event files and their corresponding bad pixel files. The obtained event files were also visually analyzed using their high energy (10--12 keV) light curves for any time periods that were highly affected by background flaring events and were then removed from the data.

We extracted the spectral files from the cleaned event files using selection criteria of FLAG==0 and PATTERN$<$=4. The source spectra were created from the  circular region of $\sim$700--1000 pixel radius, centered on the AGN position, while their corresponding background spectra were selected from the nearby-source (on the same CCD if possible) and source-free circular area having the radius of $\sim$1000 pixels. Throughout this paper, the spectra were analyzed using an X-Ray Spectral Fitting Package ({\sc xspec}) version 12.14.1\footnote{\url{https://heasarc.gsfc.nasa.gov/docs/xanadu/xspec/}} \citep{Arnaud1996}, in which all spectra were re-binned to have at least 20 counts for each individual data point so that are appropriate for applying $\chi^{2}$ minimization method during the fit. 

In case of extracting the AGN light curves, the selection criteria of \#XMMEA\_EP and PATTERN$<$=4 with same source and background regions used in the spectral extraction were also applied to extract the AGN light curves in the two energy bands, 0.3--1 keV and 1--4 keV, with a time resolution of 1.0 s for the analysis in this work. Moreover, we examine wavelet lags both on a per-observation basis and from combined datasets constructed by stitching light curves under two scenarios. In the first case, only the last 3 observations, taken within $\sim 6$~months from December 2013 -- May 2014, were combined to minimize long-term variability and ensure consistent source properties. In the second case, all 5 observations were combined to maximize the total duration. These combined observations are referred to as Fairall~9, c3 and Fairall~9, c5, respectively. In both cases, the light curves were directly stitched together. See Section 5 for simulated wavelet lags analogous to these cases.

\begin{table}
\begin{center}
   \caption{\emph{XMM-Newton} observations used in this work.} \label{tab:xmm_obs}
   \label{tab_obs}
   \begin{threeparttable}
    \begin{tabular}{lccc}
    \hline
    Obs. ID$^{a}$ & Obs. date$^{b}$ &Exp. length$^{c}$ & Useful Exp.$^{d}$ \\
            &           &  (ks)            & (ks)   \\
    \hline
0101040201	& 2000-07-05  & 29.00    & 25.83 \\
0605800401	& 2009-12-09  & 129.57   & 87.88 \\
0721110101	& 2013-12-19  & 41.25    & 24.08 \\
0721110201	& 2014-01-02  & 49.68    & 23.86 \\
0741330101	& 2014-05-09  & 124.84   & 80.96 \\
     \hline
     \end{tabular}
    \begin{tablenotes}
    \textit{Note.} $^{a}$The observation ID and $^{b}$its corresponding observation date. $^{c}$The remain exposure length after the high background flaring period(s) was removed and $^{d}$its corresponding useful exposure time obtained from the summation of good time intervals within the exposure length. (see Section~\ref{sec:Results}). 
    \end{tablenotes}
    \end{threeparttable}
    \end{center}
\end{table}

\section{Spectral model}

While a wide range of spectral models have been proposed and investigated for Fairall~9 \citep{Emmanoulopoulos2011, Lohfink2016, Liu2020, Hagen2023-Feb}, we adopt the {\sc reXcor} model \citep{Xiang2022} which attributes the soft excess to Comptonization of thermal emission in a warm corona and self-consistently incorporates both ionized relativistic reflection and warm corona emission. Indeed, this represents a scenario that has not been previously explored for this source. Essentially, the {\sc reXcor} model assumes lamp-post geometry that the hot corona locates at the height $h$ on rotational axis of the central black hole with a standard accretion disc \citep{Shakura1973}. The continuum X-ray emission from the hot corona has a power-law shape characterized by a photon index ($\Gamma$) with high energy cut-off at 300~keV. The ionization parameter at particular disc radius is defined as
\begin{equation}
\xi(r) = \frac{4\pi F_{\rm X}(r)}{n_{\rm H}(r)} \;,
    \label{eq:r_warm}
\end{equation}          
where $F_{\rm X}(r)$ and $n_{\rm H}(r)$ are the incident flux and the hydrogen number density in the slab of the disc at radius $r$, respectively. The {\sc reXcor} model adopts a disc density similar to that of \citep{Shakura1973}, consistent with observational evidence \citep{Jiang2020}.

Furthermore, {\sc reXcor} assumes the accretion disc extends from $r_{\rm in} = r_{\rm ISCO}+0.5$ to $r_{\rm out} =400~r_{\rm g}$. Note that $r_{\rm ISCO}$ is the innermost stable circular orbit and $r_{\rm g} = GM/{c^2}$, where $G$ is the gravitational constant, $M$ is the central mass, and $c$ is the speed of light. The warm corona is a Comptonizing layer with a constant density at the surface of the accretion disc between $r_{\rm in}$ and $r_{\rm warm}$. The redius $r_{\rm warm}$ is defined such that $\xi(r_{\rm warm}) = 5$ erg s cm$^{-1}$.

The {\sc reXcor} model incorporates several physical parameters that characterize the warm corona and the distribution of energy around the black hole. The parameter $f_{\rm X}$, ranging from 0.02 to 0.2, represents the fraction of the local accretion disc dissipation rate, $D(r, \lambda)$, within $r = 10\,r_{\rm g}$, that powers the compact, hot lamp-post corona. The heating fraction $h_{\rm f}$, which spans values from 0.0 to 0.8, corresponds to the fraction of the energy $D(r, \lambda)$ that is uniformly deposited in the warm corona layer, contributing to its heating. The remaining energy at each radius, given by $(1 - f_{\rm X} - h_{\rm f}) D(r, \lambda)$, is assumed to be thermalized at the base of the warm corona and emitted as blackbody radiation. The Thomson optical depth of the warm corona, $\tau$, is given in the range of 10 to 30. The warm corona reprocesses the primary X-rays from the hot corona and scatters thermal photons from the underlying disc, producing a smooth soft excess in the observable spectrum \citep{Xiang2022, Ballantyne2024}. Relativistic effects are included using the {\sc relconvlp} model \citep{Dauser2013} with an inclination angle of $30^\circ$. For full details, see \cite{Xiang2022}.

The output spectrum from {\sc reXcor} is a combination of ionized reflection from the accretion disc and the emission from the warm corona. There are several {\sc reXcor} model grids\footnote{\url{https://github.com/HEASARC/xspec_localmodels/tree/master/reXcorb}} provided to account for variations in AGN properties, including coronal height ($h$ = 5 or 20 $r_{\rm g}$), Eddington ratio ($\lambda$ = 0.01 or 0.1), photon index ($\Gamma$ = 1.5--2.5) and black hole spin ($a$ = 0.99 or 0.9). We adopt $\lambda = 0.1$, following \citet{Hagen2023-Feb}, and assume a high spin of $a = 0.99$, as required by spectral fits from {\it XMM-Newton} and {\it NuSTAR} observations \citep{Lohfink2016}.

The X-ray spectral data in the 0.3--10.0 keV band are fitted using the spectral model of
\\ [5pt]
{\sc tbabs $\times$ (cutoffpl + reXcor + xillver)},
\\ [5pt]
where {\sc tbabs} represents the Galactic absorption along the line of sight, with the hydrogen column density fixed at $n_{\rm H} = 2.85\times10^{20}~{\rm cm}^{-2}$ adopted from HEASARC N$_{\rm H}$\footnote{\url{https://heasarc.gsfc.nasa.gov/cgi-bin/Tools/w3nh/w3nh.pl}} \citep{HI4PI2016}. The {\sc cutoffpl} component describes the primary X-ray continuum from the hot corona, modeled as a power-law with high-energy exponential cutoff fixed at 300 keV following the {\sc reXcor} prescription. The {\sc reXcor} model describes the warm corona emission and relativistic disc reflection. The {\sc xillver} model \citep{Garcia2010,Garcia2013} is employed to account for the distant reflection from neutral material with the reflection fraction fixed at –1. Throughout our analysis, the photon index ($\Gamma$) of {\sc cutoffpl}, {\sc reXcor} and {\sc xillver} were tied and limited within the range of 1.5–2.5 as well as the inclination $i$ that was fixed at $30^{\circ}$, which is consistent with the available values and assumption in the {\sc reXcor} model. The normalization of every components are free to vary and other parameters not mentioned are set to default.

For some observations, we included one or more {\sc zgauss} components to improve the fit in the soft X-ray band. The line energy ($E$) was allowed to be varied within the 0.3--1 keV range, and the line width ($\sigma$) was constrained between 0.06 and 0.1 keV to capture narrow emission features contributing exclusively to the soft band, rather than the harder Fe-K region. Once the best-fitting spectral model was obtained, the corresponding flux was estimated using the {\texttt flux} command in {\sc xspec}. Previous studies have found no compelling evidence for a significant warm absorber in Fairall~9 \citep[e.g.,][]{Emmanoulopoulos2011, Patrick2011, Lohfink2012}, and thus we did not include this component in our primary model. Nonetheless, we also tested alternative fits incorporating a warm absorber component using {\sc absori} \citep{Done1992, Magdziarz1995, Zdziarski1995}.

\section{Wavelet coherence and time lags}


In Fourier analysis, the lag-frequency spectra between two energy bands can be calculated using the standard Fourier analysis method \citep{Nowak1999}. The cross-spectrum is first calculated as $C(f)=S^{*}(f)H(f)$, where $S(f)$ and $H(f)$ are the Fourier transforms of the soft and hard band light curves, respectively, and $S^{*}(f)$ denotes the complex conjugate of the soft band component. The phase difference, $\phi(f)$, is defined as the argument of the cross-spectrum $C(f)$, and the corresponding time lag is given by
\begin{equation}
\tau_{\rm lag} = \phi(f)/(2 \pi f) \;.
    \label{eq:fouroer-lag}
\end{equation}
It is important to note that the phase difference is defined in the range $-\pi$ to $+\pi$, which causes phase wrapping at high frequencies (i.e., when $f \geq 1/2\tau$), where the lag-frequency profile oscillates around zero.

Moreover, the coherence (i.e., power of correlation) can be calculated by 
\begin{equation}
\gamma^2 = \frac{{\left| \left\langle S^{*}(f)H(f) \right\rangle \right|}^{2}}{{\left\langle \left| S^{2}(f) \right| \right\rangle}{\left\langle \left| H^{2}(f) \right| \right\rangle}} \;,
   \label{eq:fourier-coherence}
\end{equation}
where angle brackets denote the average of each quantity across the frequency components within each bin. The uncertainty in time-lag estimates increases as the coherence decreases, so the lags are reliable only when the coherence between two energy bands is high \citep{Epitropakis2016a, Epitropakis2017}.




Meanwhile, the wavelet transform is a time-frequency decomposition technique that enables the analysis of non-stationary signals by preserving both temporal and spectral information. Unlike the Fourier transform which provides a time-averaged frequency representation in terms of amplitude and phase, the wavelet transform captures localized spectral features by convolving the signal with a set of scaled and shifted wavelets. This enables detection of time-dependent variability across a wide range of frequency scales \citep{Ghosh2023, Wilkins2023}. A wavelet transform of $x(t)$ can be written as:
\begin{equation}
T(a,b)= w(a)\int_{-\infty}^{+\infty}{x(t)~\psi^{*}\left( \frac{t-b}{a} \right)dt} \;,
    \label{eq:wavelet-lag}
\end{equation}
where $w(a)$ is a weighting function, and the parameters $a$ and $b$ represent the scale and translation, respectively \citep{Addison2017}. The scale parameter $a$ controls the dilation or compression of the wavelet basis function $\psi(t)$, analogous to frequency in the Fourier transform. The translation parameter $b$ shifts the wavelet along the time axis, allowing localization of signal features in time.

An important consideration in wavelet analysis is the cone of influence (COI), which indicates the region of the time-frequency plane where edge effects become significant due to the finite length of the input light curves. When the wavelet overlaps the edges of the data, i.e. start or end of the time series, it lacks sufficient surrounding data to fully capture the frequency content. Consequently, meaningful analysis should be restricted to within the COI to ensure the robustness of the results \citep{Torrence1998, Ghosh2023}. 

Similarly to the Fourier domain, coherence can also be defined in a wavelet analysis, referred to as wavelet coherence. From this, phase lags and time lags can be computed in a wavelet analysis. While Fourier coherence produces a coherence spectrum averaged over the entire time series, wavelet coherence can capture both the strength and the phase relationship between signals as they evolve \citep{Wilkins2023}. This enables detection of transient and frequency-dependent coupling that might remain hidden in global Fourier-based analyses. Here, we calculate wavelet coherence spectra of the individual Fairall~9 observation as well as the stitching light curves, c3 and c5, using the \texttt{wavelet\_coherence} function from the Python package \texttt{Pyleoclim} \citep{Khider2022a, Khider2022b}\footnote{\url{https://github.com/LinkedEarth/Pyleoclim_util}}, which is designed for wavelet analysis. The significance are calculated using \texttt{coh.signif\_test} function, based on Monte Carlo simulations with phase-randomized surrogates. Note that for the wavelet analysis, the light curves were subsequently re-binned to reduce computational time. The individual light curves were binned at 50 s intervals, while longer combined light curves were binned at 100 s intervals. Besides measuring coherence amplitude, phase lag directions that show the relative timing between variations in the two signals are also calculated. Using this phase information, we examine how the phase relationship between soft and hard X-ray light curves changes over time and frequency. The phase coherence helps analyze localized correlations and identify when one band leads or lags the other, and on what timescales.

\section{Simulation tests on wavelet lags}

While combined stitching light curves is known to introduce potential discontinuities that can affect Fourier-based lag spectra, the wavelet framework is inherently more localized in both time and frequency and is therefore more robust to such effects. For this reason, we do not attempt to compute Fourier lag spectra from stitched light curves. Fourier analyses restricted to individual observations are further limited by their short duration, which prevents access to the low-frequency range of interest. Our primary focus then remains on wavelet-based lags, which are better suited to capture transient lag signatures and track their temporal evolution.

To test whether stitching introduces bias (e.g. spurious features) in the wavelet lags, we simulate AGN light curves under controlled conditions. Five hard-band light curves of lengths 25000, 50000, 75000, 100000, and 125000~s were generated from a red-noise power spectral density with a power-law index of 2 using the \texttt{stingray.simulator} package \citep{Huppenkothen2019, Bachetti2024}. These simulations assume a stationary power spectrum, meaning that statistical properties of the light curves, including their power spectral density (PSD), remain constant throughout their entire duration. This simplifying assumption allows us to generate light curves with controlled properties, enabling us to isolate the specific effects of stitching on wavelet lag estimates. The corresponding soft-band light curve were then simulated with a fixed delay of 1000 s relative to each hard-band curve. The wavelet lags between the hard- and soft-band light curves are analyzed both in their original, unstitched form and after direct concatenation of the segments.

\begin{figure*}
    \centering
    \includegraphics[width=\linewidth]{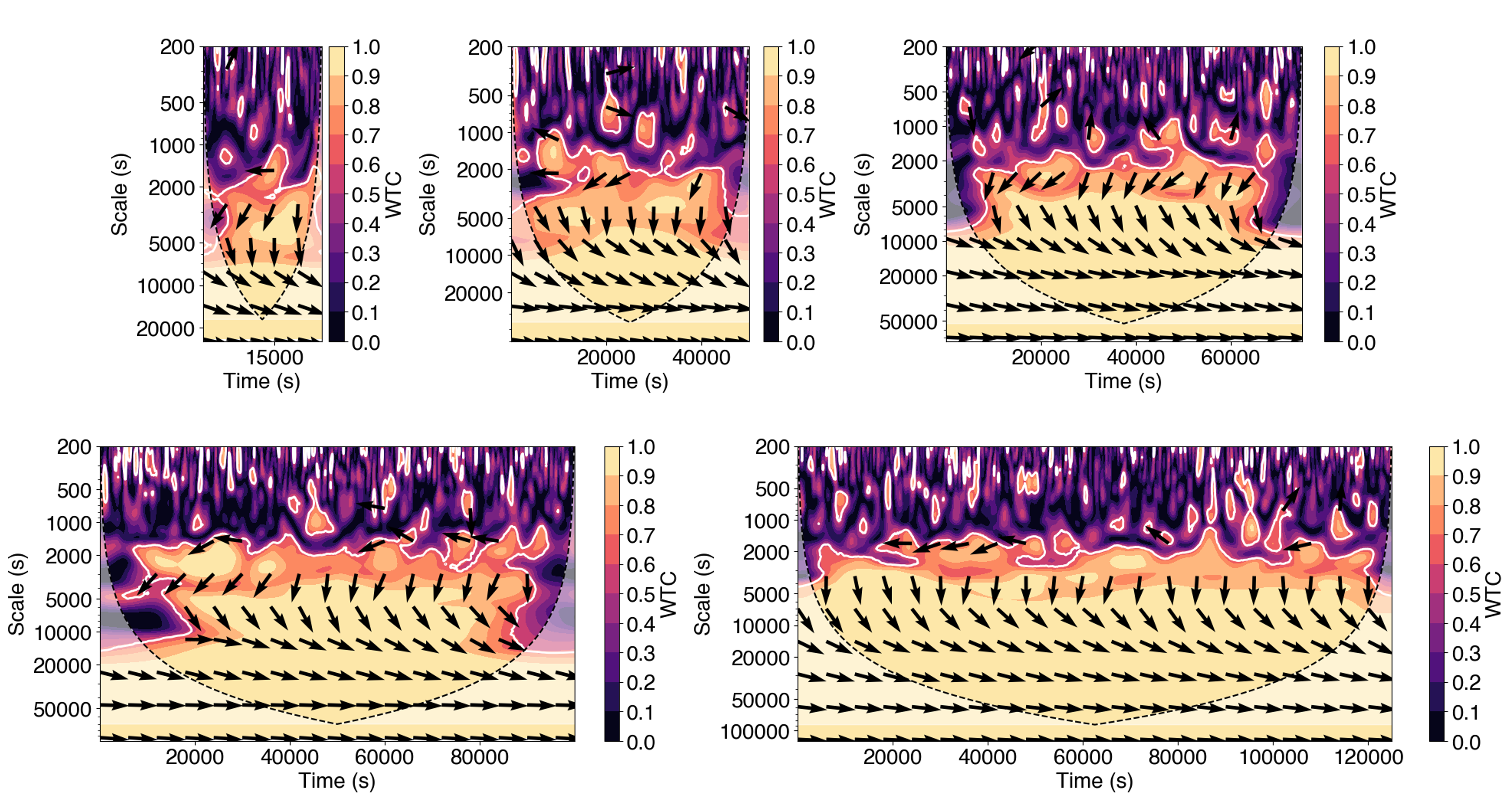}
    \caption{Simulated wavelet coherence (WTC) spectra for various cases of light curves of lengths 25000, 50000, 75000, 100000, and 125000 s, each mimicking a single observation. In all cases, the soft-band light curves are delayed by 1000 s relative to the hard-band light curves. The color scale shows wavelet coherence magnitude, where warmer colors indicate higher coherence between the energy bands with 90 per cent confidence interval included with white contour line. The y-axis shows timescales, which are the inverse of temporal frequency. Arrows indicate phase lag direction: right-pointing arrows show in-phase variations, left-pointing show anti-phase, downward arrows indicate soft lags, and upward arrows show hard lags. The dashed line marks the cone of influence (COI), beyond which edge effects may compromise signal reliability.}
    \label{fig-coh-sim}
\end{figure*}

Fig.~\ref{fig-coh-sim} shows the simulated wavelet coherence (WTC) spectra for each individual case, where the soft-band light curve is delayed by 1000 s relative to the hard-band light curve. The color map shows the strength of coherence, with values $\gtrsim 0.8$ (light yellow regions) indicating strong correlation between the bands. We also show contour lines indicating the 90\% confidence intervals in the plots. High coherence is observed at timescales of $\gtrsim 2000$~s across all WTC spectra, corresponding to frequencies of about $5\times10^{-4}$ Hz. At lower timescales, phase wrapping is evident. However, longer timescales lead to a narrower area of COI, meaning the inferred results are reliable only in shorter light curve segments. The overplotted arrows represent the phase difference between signals, where rightward arrows indicate in-phase behavior and downward- (upward-) pointing arrows denote phase lags corresponding to soft (hard) lags. At long timescales, the arrows predominantly point down and to the right within high-coherence regions, indicating in-phase variations where the soft band lags the hard band. This pattern persists toward shorter timescales until noise or finite sampling effects dominate, and it is consistent across all simulated cases, as they were generated under identical conditions.

\begin{figure*}
    \centerline{
        \includegraphics[width=\textwidth]{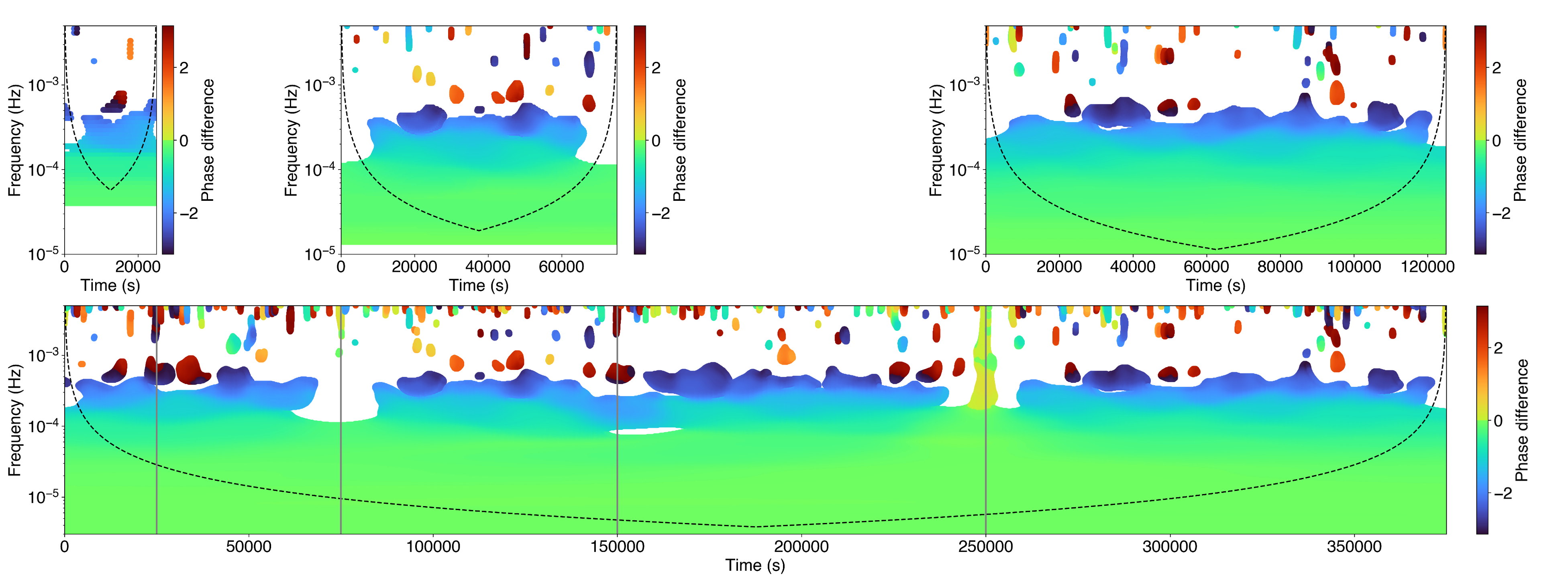}
    }
    \centerline{
        \includegraphics[width=\textwidth]{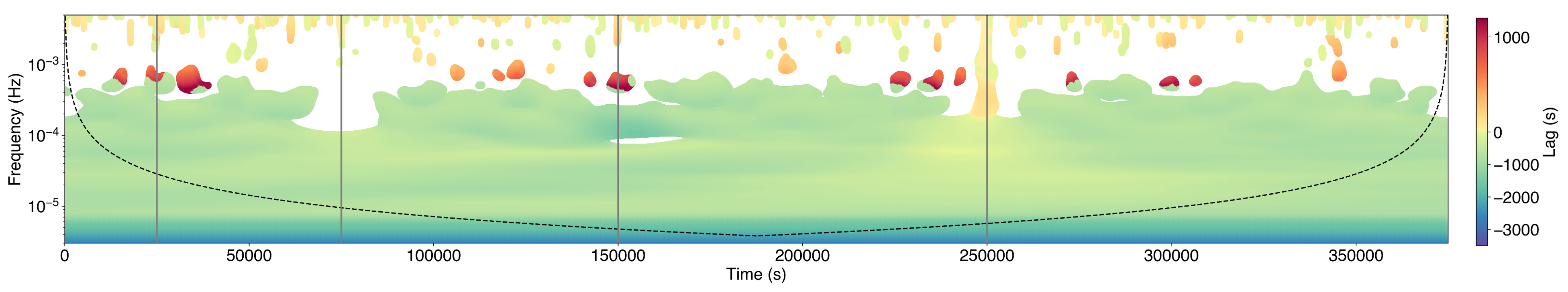}
    }
    \caption{Time-resolved phase differences for individual cases (upper panel) and combined, stitched cases (middle panel) corresponding to the simulated WTC shown in Fig.~\ref{fig-coh-sim}. The vertical solid lines in gray indicate the boundaries between individual observations. The color scale represents phase differences: negative values (green–blue–black) indicate soft lags, where the soft band lags behind the hard band, while positive values (yellow–orange–red) indicate hard lags. The bottom panels show the corresponding time-lag profiles, with color denoting lag amplitude. Regions with signal coherence below 0.7 are excluded. The wavelet phase and lag spectra of the stitched light curves successfully recover the imposed delays across the relevant frequencies, consistent with the results from individual segments. See text for more details.}
    \label{fig-phase-dif-sim}
\end{figure*}

Fig.~\ref{fig-phase-dif-sim} (top and middle panels) shows the phase shift between the cases of using the individual light curves and combined stitching light curves. For clarity, only three individual cases are displayed. We used a coherence threshold to examine the phase structure throughout the time-frequency domain. We set the coherence threshold to $\geq 0.7$ to focus on the most reliable, highly coherent regions, which are also likely to be significant because they fall within the 90\% confidence intervals shown in Fig.~\ref{fig-coh-sim}. The areas inside the cone of influence (COI) are broader when using stitched light curves; however, the overall phase-difference patterns remain consistent with those derived from individual light curves with only small differences appear near the boundaries between segments, indicating that the effects of stitching are minor.

\begin{figure*}
    \centering
    \includegraphics[width=\textwidth]{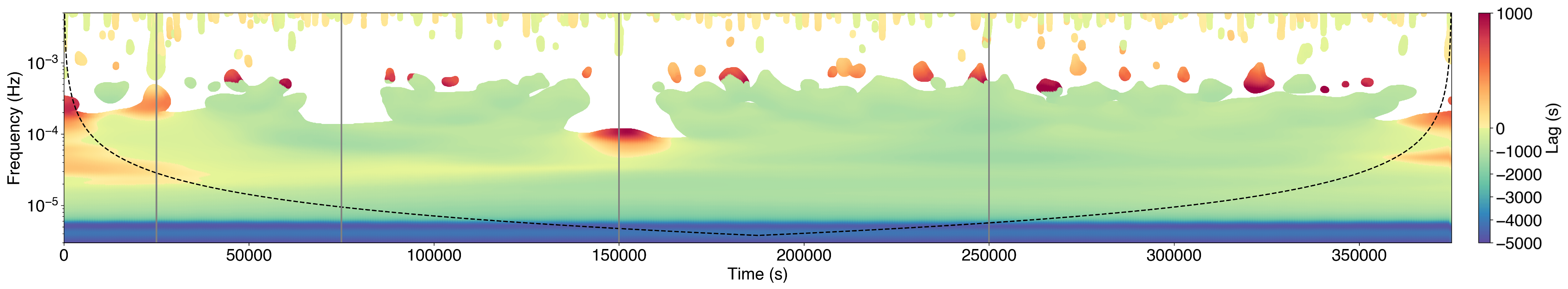}
    \caption{Wavelet lag profile as same as Fig.~\ref{fig-phase-dif-sim} (bottom panel), but with Poisson noise included.}
    \label{fig-lag-sim-noise}
\end{figure*}

The corresponding time-lag profiles are shown in Fig.~\ref{fig-phase-dif-sim} (bottom panel). Soft lags of about 1000~s are evident at low frequencies, as expected for a lag spectrum from a primary light curve with an identical curve delayed by 1000~s. As the individual light curves were simulated with a constant time delay, the lag is expected to remain approximately constant across all frequencies, except where phase wrapping is introduced by the sampling method. Importantly, the stitched light curves reproduced the same lag–frequency patterns without introducing significant artificial structures. Meanwhile, Poisson noise degrades the quality of the results (Fig.~\ref{fig-lag-sim-noise}); however, a lag amplitude of about 1000 s can still be recovered consistently across the relevant frequencies and light-curve segments. As an example, we also test the stitching method on light curves with different mean count rates of 3, 6, and 9 counts/s, keeping all other parameters fixed. We find that stitching preserves wavelet coherence profiles regardless of mean count rate differences in individual segments. This demonstrates that the wavelet techniques can be a useful strategy for probing time-lag signatures at lower frequencies than those accessible with a single light curve using the traditional lag–frequency spectrum.


\section{Fairall~9 Results}
\label{sec:Results}

\begin{figure*}
  \centering

  \begin{subfigure}{0.325\textwidth}
    \includegraphics[width=\linewidth]{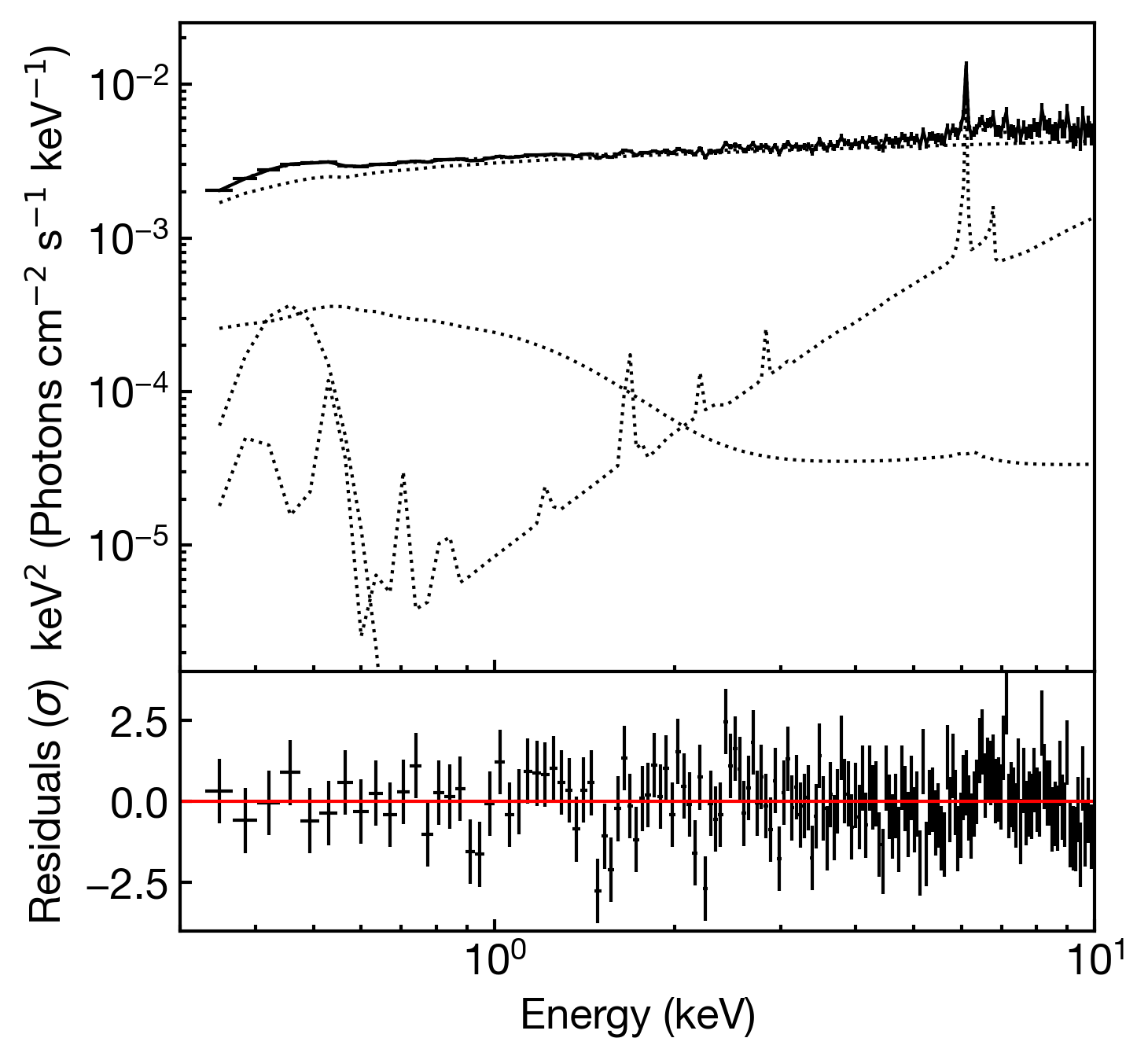}
  \end{subfigure}
  \begin{subfigure}{0.325\textwidth}
    \includegraphics[width=\linewidth]{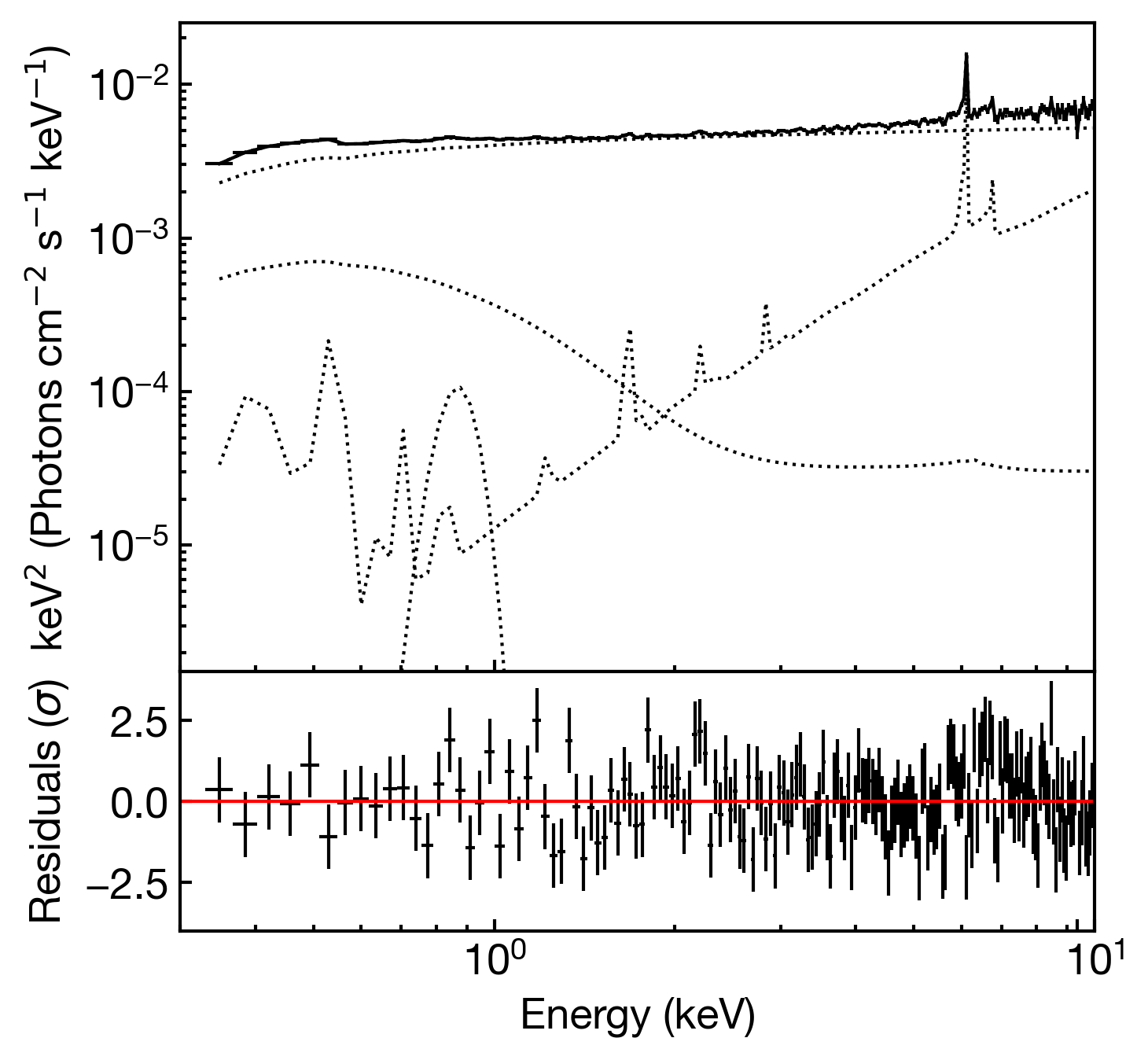}
  \end{subfigure}
  \begin{subfigure}{0.325\textwidth}
    \includegraphics[width=\linewidth]{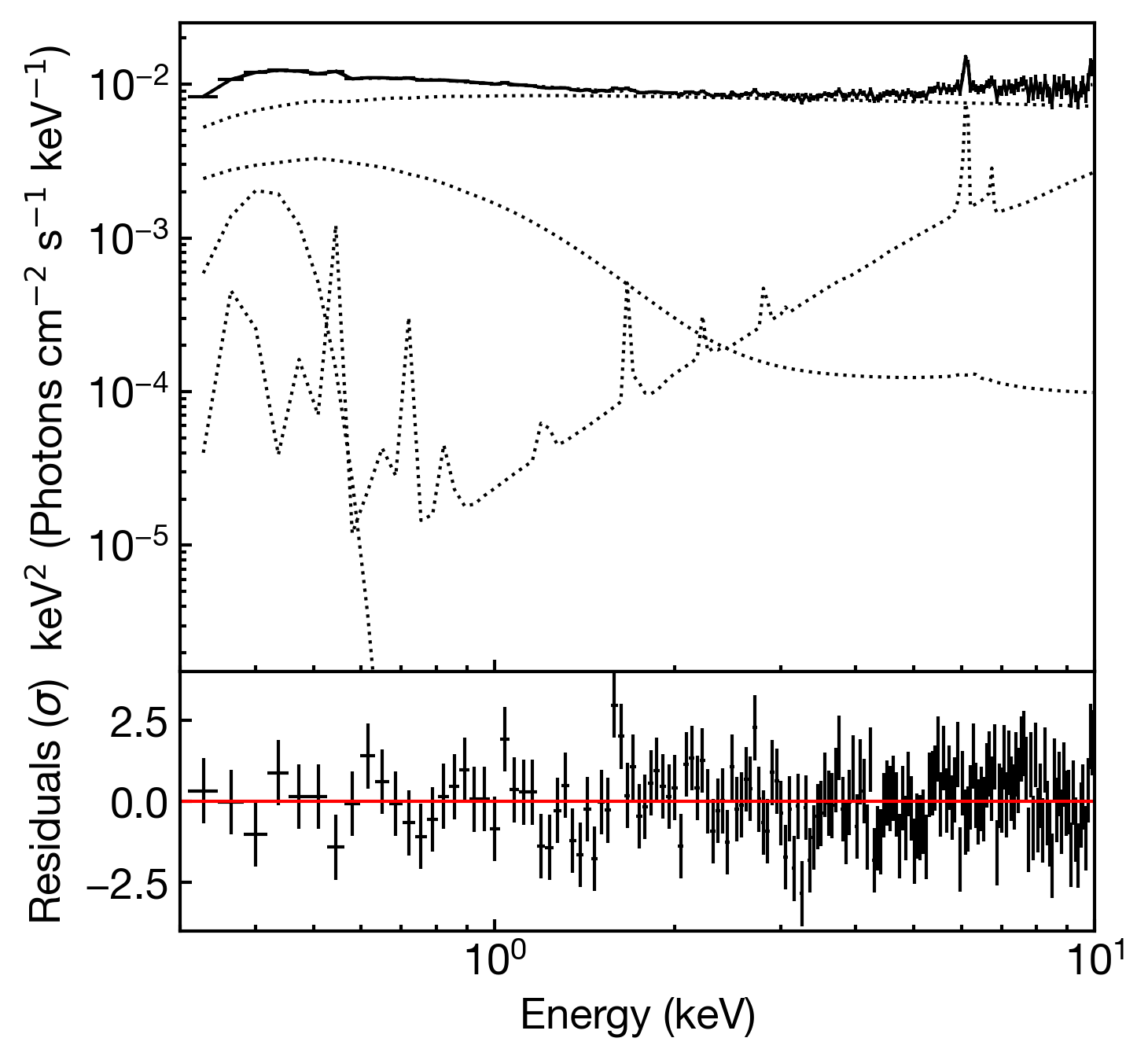}
  \end{subfigure}

  \begin{subfigure}{0.325\textwidth}
    \includegraphics[width=\linewidth]{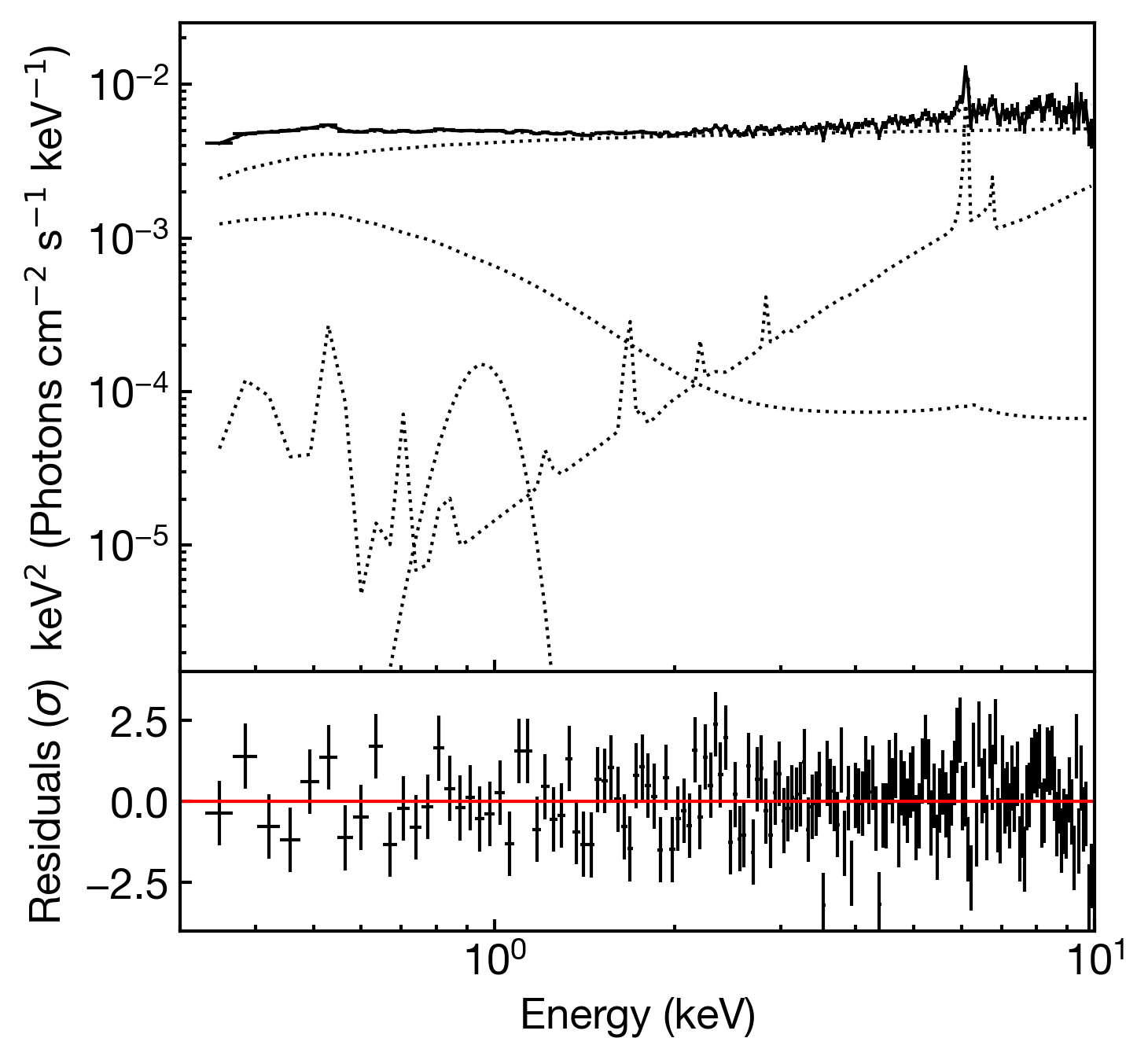}
  \end{subfigure}
  \begin{subfigure}{0.325\textwidth}
    \includegraphics[width=\linewidth]{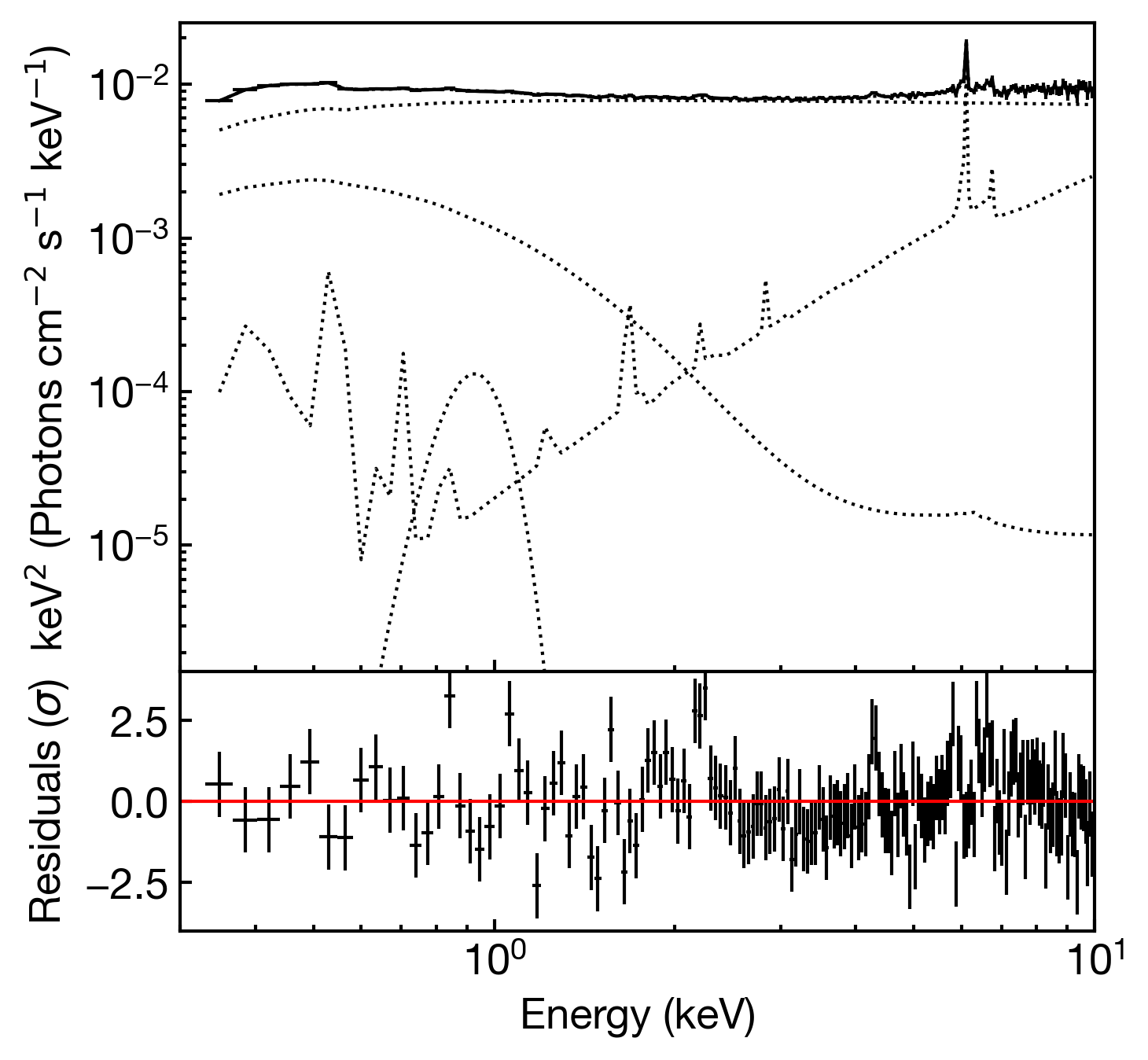}
  \end{subfigure}

  \caption{Spectral data and residuals (black points) relative to the best-fitting model with $h=5~r_{\rm g}$. The panels, arranged from top-left to bottom-right, correspond to the {\it XMM-Newton} observations in chronological order, from the earliest (Obs. ID 0101040201) to the latest (Obs. ID 0741330101). Individual model components are shown as dotted lines in each panel. Note that the residuals are presented in the unit of $\sigma$}
  \label{fig-eeuf-h5}
\end{figure*}

\begin{figure}
    \centering
    \includegraphics[scale=0.65]{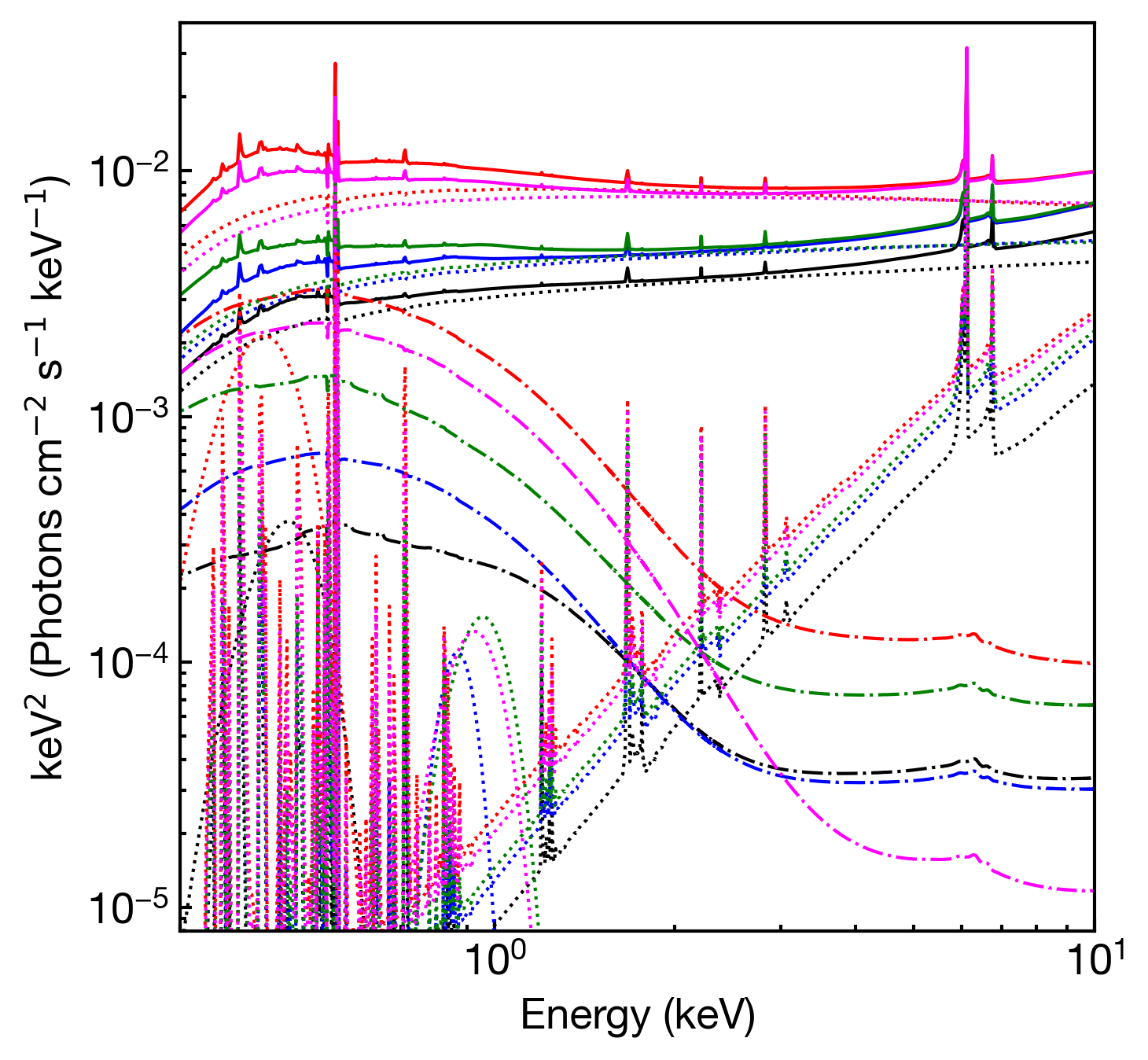}
    \caption{Comparison of spectral model components from the best-fitting model with $h = 5~r_{\rm g}$. The total model is shown as solid lines, the {\sc reXcor} component as dot-dashed lines, and other components ({\sc cutoffpl}, {\sc xillver}, and additional {\sc zgauss}) as dotted lines. Black, blue, red, green, and magenta correspond to different observations, ordered chronologically from the earliest (Obs. ID 0101040201) to the latest (Obs. ID 0741330101).}
    \label{fig-eemo-h5}
\end{figure}

The best-fitting spectral parameters of Fairall~9 are summarized in Table~\ref{tab:bestfit_a099_l01}, with representative fitting results for the $h = 5~r_{\rm g}$ case illustrated, as an example, in Fig.~\ref{fig-eeuf-h5}. Most observations are reasonably well fitted by either configuration of $h = 5~r_{\rm g}$ or $20~r_{\rm g}$, with reduced chi-squared values $\chi^{2}_\nu \lesssim 1.5$. The photon index remains consistently within the range of 1.86--2.05. The {\sc reXcor} model parameters (i.e., $f_{\rm X}$, $h_{\rm f}$, and $\tau$) vary across the observations, and we do not observe any clear correlation among them. However, focusing on the energy distribution between the hot and warm corona,  the results suggest the consistent presence of a warm corona ($h_{\rm f} \ne 0$), with a larger fraction of the accretion energy dissipated in the warm layer ($h_{\rm f} > f_{\rm X}$). The optical depth of the warm corona varies between observations. Additionally, the Fe abundance derived from the {\sc xillver} distant reflection is found to be sub-solar, with $A_{\rm Fe} < 1$.

In Fig.~\ref{fig-eemo-h5}, the {\sc reXcor} components in the soft X-ray band ($<$1 keV), shown as dot-dashed lines, follow the same trend as the total fluxes, with lower-flux observations exhibiting weaker soft excess features. To account for unresolved narrow emission lines commonly observed in AGN spectra, one or two additional {\sc zgauss} components are included in the model. Most observations require the inclusion of a narrow line near 0.41–0.46 keV, likely associated with nitrogen transitions such as {\sc N VI} ($\sim$0.50 keV) or {\sc N VII} ($\sim$0.45 keV) \citep{Ballantyne2024}. As discussed in \citet{Xiang2022}, although the {\sc reXcor} model includes emission from {\sc N VII}, it appears to underestimate the strength of this feature in some spectra, likely due to the high sensitivity of He-like transitions to the property of plasma. In some observations, though less frequently, the fit was also improved by including narrow lines at approximately 0.34–0.39 keV and/or 0.90–0.99 keV. The lower-energy feature may correspond to {\sc C VI} emission, reported at $\sim$0.37 keV by \citet{RossFabian2005}.

Additional tests are performed to investigate the presence of both warm and additinal neutral absorption. When including a warm absorber component ({\sc absori}), the hydrogen column density ($n_{\rm H}$) is significantly low and the ionization parameter ($\xi$) remains unconstrained. This indicates that a warm absorber does not significantly improve the fit and is not statistically required. Similarly, testing for additional neutral absorption using a second {\sc phabs} component, beyond the fixed line-of-sight hydrogen column density of $n_{\rm H} = 2.85 \times 10^{20}~{\rm cm}^{-2}$, yielded no significant improvement, and thus this component is excluded from the final model.

\begin{table*}
 \caption{The best-fitted results for each observation. Note that we use {\sc reXcor} grid models of black hole spin $a$ = 0.99 and Eddington ratio $\lambda$ = 0.1, and investigate both coronal heights of 5 and 20 $r_{\rm g}$. The photon index ($\Gamma$) of {\sc cutoffpl}, {\sc reXcor} and {\sc xillver} were tied together. A symbol ‘$p$’ in the error bar indicates that a parameter were pegging at the upper or lower limit of the model. If the parameter has unconstrained errors on both ends, it is indicated with a ‘$\ast$’. The 0.3–10 keV fluxes are included as $F$ which is in the unit of erg cm$^{-2}$ s$^{-1}$).}
 \label{tab:bestfit_a099_l01}
 \begin{tabularx}{\textwidth}{lp{60pt}XXXXX}
 
  \hline \\ [\dimexpr-\normalbaselineskip+2pt]
  
        & & \multicolumn{5}{c}{Observation ID} \\ [5pt]
        & & 0101040201 & 0605800401 & 0721110101 & 0721110201 & 0741330101 \\ [5pt]
        
  \hline \\ [2pt]

  
  
  {\sc cutoffpl}    & $\Gamma$                  & $1.87_{-0.3}^{+0.06}$ & $1.90_{-0.04}^{+0.02}$ & $2.08_{-0.05}^{+0.03}$ & $1.92_{-0.1}^{+0.07}$ & $2.03_{-0.02}^{+0.02}$ \\ [5pt]
                    & norm ($\times 10^{-3}$) & $3.24_{-0.9}^{+0.2}$ & $4.22_{-0.1}^{+0.08}$ & $8.85_{-0.5}^{+0.3}$ & $4.40_{-0.5}^{+0.2}$ & $8.11_{-0.3}^{+0.2}$ \\ [5pt]

  {\sc reXcor}      & $f_{\rm X}$               & $0.09_{-0.02}^{+0.11p}$ & $0.20_{-0.14}^{+0p}$ & $0.16_{-0.1}^{+0.04p}$ & $0.10_{-0.06}^{+0.1p}$ & $0.04_{-0.02p}^{+0.1}$ \\ [5pt]
 ($h$ = 5 $r_{\rm g}$)   & $h_{\rm f}$          & $0.21_{-0.21p}^{+0.4}$ & $0.56_{-0.3}^{+0.24p}$ & $0.49_{-0.1}^{+0.31p}$ & $0.40_{-0.2}^{+0.4}$ & $0.71_{-0.2}^{+0.09p}$ \\ [5pt]
                    & $\tau$                    & $10.0_{-0p}^{+11.9}$ & $25.1_{-10.3}^{+4.9p}$ & $19.1_{-4.7}^{+10.9p}$ & $15.0_{-2.9}^{+15.0p}$ & $23.2_{-7.0}^{+6.8p}$ \\ [5pt]
                    & norm ($\times 10^{-17}$) & $1.38_{-1.2}^{+0.8}$ & $0.50_{-0.06}^{+0.4}$ & $3.36_{-0.7}^{+1.7}$ & $2.51_{-1.8}^{+3.0}$ & $1.70_{-0.5}^{+1.4}$ \\ [5pt]
  
  {\sc xillver}     & $A_{\rm Fe}$              & $0.79_{-0.2}^{+9.19p}$ & $0.65_{-0.1}^{+0.1}$ & $0.50_{-0p}^{+0.2}$ & $0.62_{-0.12p}^{+0.6}$ & $0.55_{-0.05p}^{+0.1}$ \\ [5pt] 
                    & norm ($\times 10^{-4}$) & $0.82_{-0.5}^{+0.2}$ & $1.20_{-0.1}^{+0.1}$ & $1.61_{-0.4}^{+0.3}$ & $1.25_{-0.5}^{+0.3}$ & $1.47_{-0.2}^{+0.2}$ \\ [5pt]

  {\sc zgauss1}     & $E_{\rm 1}$ (keV)         & $0.45_{-0.06}^{+0.02}$ & $0.42_{-0.05}^{+0.02}$ & $0.41_{-0.03}^{+0.007}$ & $0.34_{-0.04}^{+0.06}$ & $0.39_{-0.05}^{+0.02}$ \\ [5pt]
                    & $\sigma_{\rm 1}$ (keV)    & $0.06^{\ast}$ & $0.08^{\ast}$ & $0.06_{-0p}^{+0.02}$ & $0.10^{\ast}$ & $0.08^{\ast}$ \\ [5pt]
                    & norm ($\times 10^{-3}$) & $0.37_{-0.1}^{+0.5}$ & $0.60_{-0.2}^{+0.5}$ & $2.76_{-0.4}^{+1.3}$ & $1.46_{-1.2}^{-0.9}$ & $2.15_{-0.9}^{+1.7}$ \\ [5pt]

  {\sc zgauss2}     & $E_{\rm 2}$ (keV)         & - & $0.90_{-0.04}^{+0.03}$ & - & $0.98_{-0.08}^{+0.02p}$ & $0.95_{-0.06}^{+0.05p}$ \\ [5pt]
                    & $\sigma_{\rm 2}$ (keV)    & - & $0.06^{\ast}$ & - & $0.10^{\ast}$ & $0.10^{\ast}$ \\ [5pt]
                    & norm ($\times 10^{-5}$) & - & $2.35_{-1.2}^{+2.8}$ & - & $4.53_{-2.8}^{+3.9}$ & $4.21_{-0.2}^{+0.2}$ \\ [5pt]

  {flux}            & $F$ ($\times 10^{-11}$)           & 2.09 & 2.73 & 5.44 & 2.96 & 4.93 \\ [5pt]
     
  \hline \\ [\dimexpr-\normalbaselineskip+2pt]

                    & {$\chi^2 / {\rm dof}$}    & 154/158 & 215/162 & 170/163 & 198/157 & 232/162 \\ [5pt]

  
  \hline \\ [2pt]
  
  
  {\sc cutoffpl}    & $\Gamma$                  & $1.87_{-0.3}^{+0.04}$ & $1.88_{-0.02}^{+0.02}$ & $2.02_{-0.07}^{+0.04}$ & $1.88_{-0.06}^{+0.07}$ & $2.06_{-0.02}^{+0.007}$ \\ [5pt]
                    & norm ($\times 10^{-3}$) & $3.21_{-0.6}^{+0.1}$ & $4.16_{-0.09}^{+0.07}$ & $8.30_{-0.6}^{+0.4}$ & $4.15_{-0.2}^{+0.1}$ & $8.05_{-0.2}^{+0.07}$ \\ [5pt]
  
  {\sc reXcor}      & $f_{\rm X}$               & $0.12_{-0.03}^{+0.08p}$ & $0.02_{-0p}^{+0.1}$ & $0.04_{-0.02p}^{+0.01}$ & $0.20_{-0.03}^{+0p}$ & $0.19_{-0.08}^{+0.01p}$ \\ [5pt]
 ($h$ = 20 $r_{\rm g}$)   & $h_{\rm f}$          & $0.14_{-0.14p}^{+0.5}$ & $0.80_{-0.3}^{+0p}$ & $0.50_{-0.04}^{+0.05}$ & $0.34_{-0.34p}^{+0.3}$ & $0.02_{-0.02p}^{+0.09}$ \\ [5pt]
                    & $\tau$                    & $10.1_{-0.1p}^{+10.9}$ & $28.2_{-5.5}^{+1.8p}$ & $13.3_{-1.3}^{+4.0}$ & $16.0_{-6.0p}^{+1.9}$ & $10.6_{-0.6p}^{+1.4}$ \\ [5pt]
                    & norm ($\times 10^{-17}$) & $0.84_{-0.7}^{+0.9}$ & $0.44_{-0.06}^{+0.1}$ & $5.49_{-2.1}^{+1.0}$ & $1.25_{-0.2}^{+4.4}$ & $3.26_{-0.2}^{+2.3}$ \\ [5pt]
                    
  {\sc xillver}     & $A_{\rm Fe}$              & $0.79_{-0.1}^{+9.21p}$ & $0.69_{-0.1}^{+0.1}$ & $0.54_{-0.04p}^{+0.3}$ & $0.72_{-0.2}^{+0.3}$ & $0.50_{-0p}^{+0.05}$ \\ [5pt] 
                    & norm ($\times 10^{-4}$) & $0.78_{-0.6}^{+0.2}$ & {$1.10_{-0.1}^{+1.1}$} & $1.27_{-0.4}^{+0.3}$ & $1.01_{-0.3}^{+0.2}$ & $1.46_{-0.2}^{+0.1}$ \\ [5pt]

  {\sc zgauss1}     & $E_{\rm 1}$ (keV)         & $0.46_{-0.04}^{+0.02}$ & $0.42_{-0.05}^{+0.02}$ & $0.41_{-0.01}^{+0.007}$ & $0.36_{-0.06p}^{+0.02}$ & $0.30_{-0p}^{+0.04}$ \\ [5pt]
                    & $\sigma_{\rm 1}$ (keV)    & $0.06^{\ast}$ & $0.07_{-0.01p}^{+0.03}$ & $0.06_{-0p}^{+0.01}$ & $0.06_{-0p}^{+0.03}$ & $0.06_{-0p}^{+0.01}$ \\ [5pt]
                    & norm ($\times 10^{-3}$) & $0.41_{-0.2}^{+0.5}$ & {$0.54_{-0.2}^{+0.4}$} & $2.49_{-0.3}^{+0.5}$ & $1.66_{-0.4}^{+2.7}$ & $2.24_{-1.3}^{+0.7}$ \\ [5pt]

  {\sc zgauss2}     & $E_{\rm 2}$ (keV)         & - & - & $0.92_{-0.06}^{+0.06}$ & $0.52_{-0.03}^{+0.02}$ & $0.43_{-0.007}^{+0.004}$ \\ [5pt]
                    & $\sigma_{\rm 2}$ (keV)    & - & - & $0.10^{\ast}$ & $0.06_{-0p}^{+0.02}$ & $0.06_{-0p}^{+0.05}$ \\ [5pt]
                    & norm ($\times 10^{-3}$) & - & - & $0.09_{-0.04}^{+0.05}$ & $0.34_{-0.1}^{+0.2}$ & $2.19_{-0.07}^{+0.02}$ \\ [5pt]

  {flux}            & $F$ ($\times 10^{-11}$)            & 2.09 & 2.73 & 5.45 & 2.96 & 4.93 \\ [5pt]
     
  \hline \\ [\dimexpr-\normalbaselineskip+2pt]

                    & {$\chi^2 / {\rm dof}$}    & 153/158 & 231/165 & 165/160 & 198/157 & 229/162 \\ [5pt]
  
  \hline
  
  \end{tabularx}
\end{table*}

\begin{figure*}
    \centering
    \includegraphics[width=\linewidth]{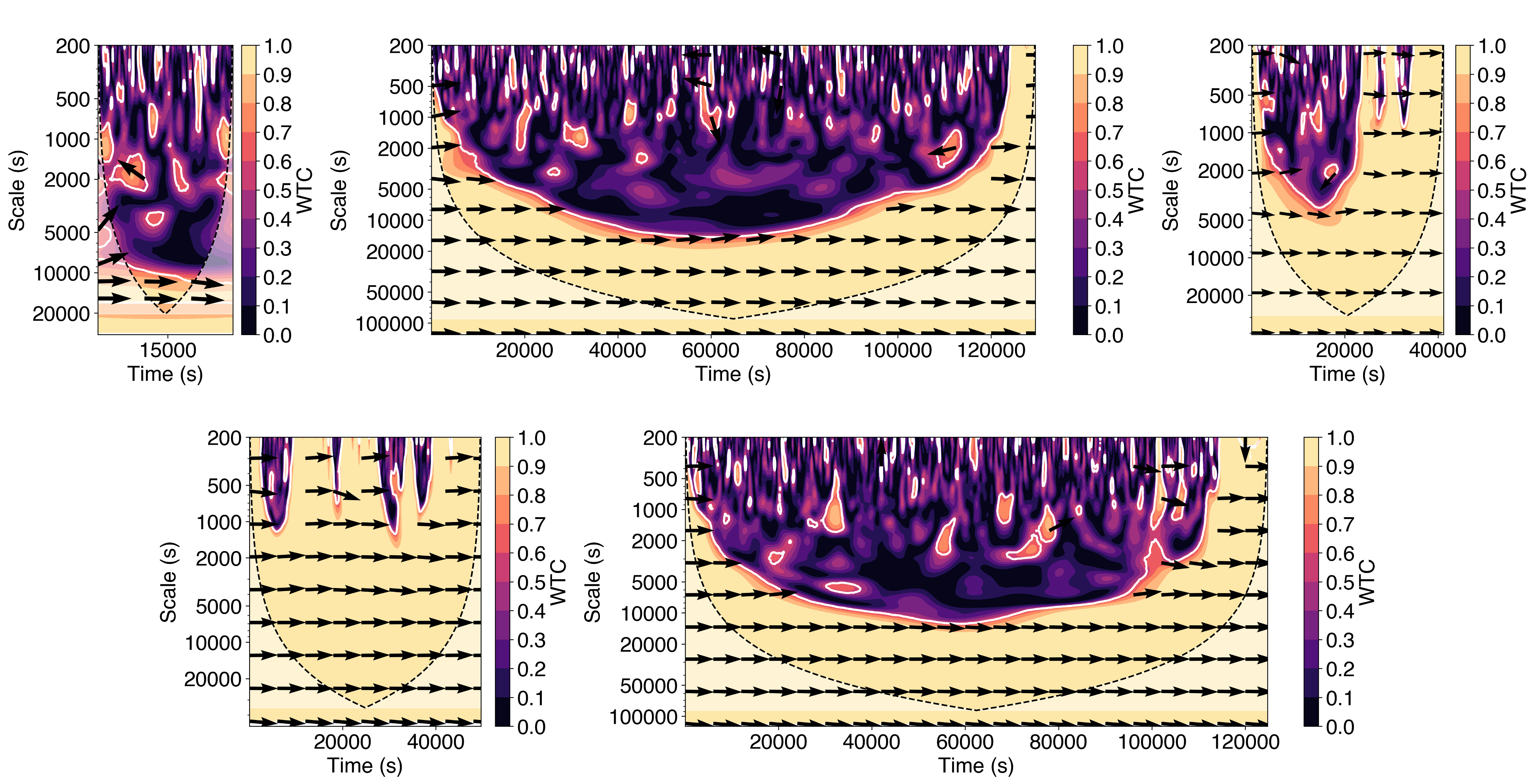}
    \caption{WTC spectra between the 0.3--1 keV and 1--4 keV bands for individual Fairall~9 observations ordered chronologically from the earliest (Obs. ID 0101040201) to the latest (Obs. ID 0741330101). Plot details follow those described in the caption of Fig.~\ref{fig-coh-sim}.}
    \label{fig-coh-F9}
\end{figure*}

To complement the spectral analysis, we employ wavelet techniques to investigate the time-dependent X-ray variability of Fairall~9. Unlike traditional Fourier methods, wavelet analysis can probe variability simultaneously in both the temporal and frequency domains. Fig.~\ref{fig-coh-F9} shows the WTC spectra between the 0.3--1 keV and 1--4 keV light curves for each observation of Fairall~9. Regions outside the COI, where edge effects become significant, are enclosed by a dashed contour that depends on the light-curve length. Overall, high coherence persists across epochs at long timescales, while lower coherence appears at shorter timescales, reflecting the stochastic nature of the variability and the occurrence of phase wrapping. In regions of high coherence, we observe minimal arrow orientation, suggesting that phase lags are small, whether negative or positive. This is in contrast with the case of IRAS~13224--3809 \citep{Wilkins2023} where overall less coherence but much stronger phase differences were observed.

\begin{figure*}
    \centering
    \includegraphics[height=0.26\textwidth]{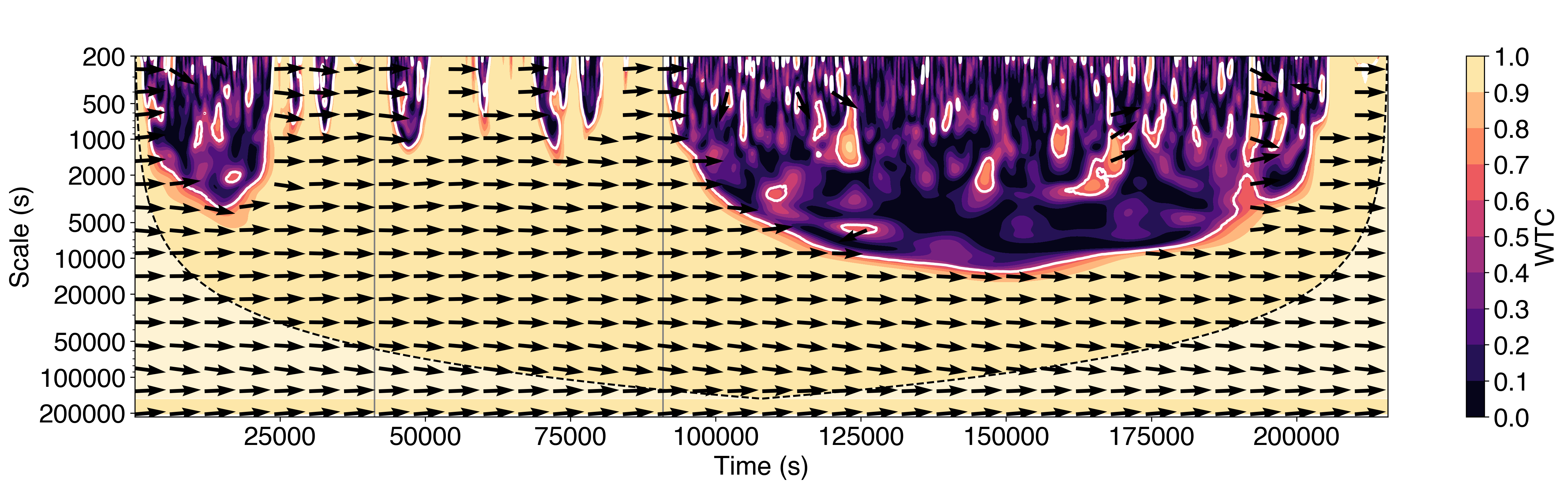}
    \caption{WTC spectra between the 0.3--1 keV and 1--4 keV bands for Fairall~9, c3.}
    \label{fig-wavelet-coh}
\end{figure*}

Fig.~\ref{fig-wavelet-coh} shows the WTC of Fairall~9, c3, in which the combined observations were obtained within a span of $\sim 6$~months. High coherence is observed primarily at timescales of $\sim 20,000$--200,000~s across most of the light curve interval. These intervals correspond to frequency ranges of $\sim 5 \times 10^{-6}$--$5 \times 10^{-5}$~Hz, which are low enough to probe reverberation lags in this source. The overall WTC structure at shorter timescales observed in individual observations is also evident in Fairall9, c3, which is expected since all these segments probe similar variability. At longer timescales, which are inaccessible in the individual observations, the wavelet analysis reveals additional high-coherence patterns, demonstrating its ability to capture variability over extended temporal ranges.

\begin{figure*}
    \centerline{
        \includegraphics[height=0.25\textwidth]{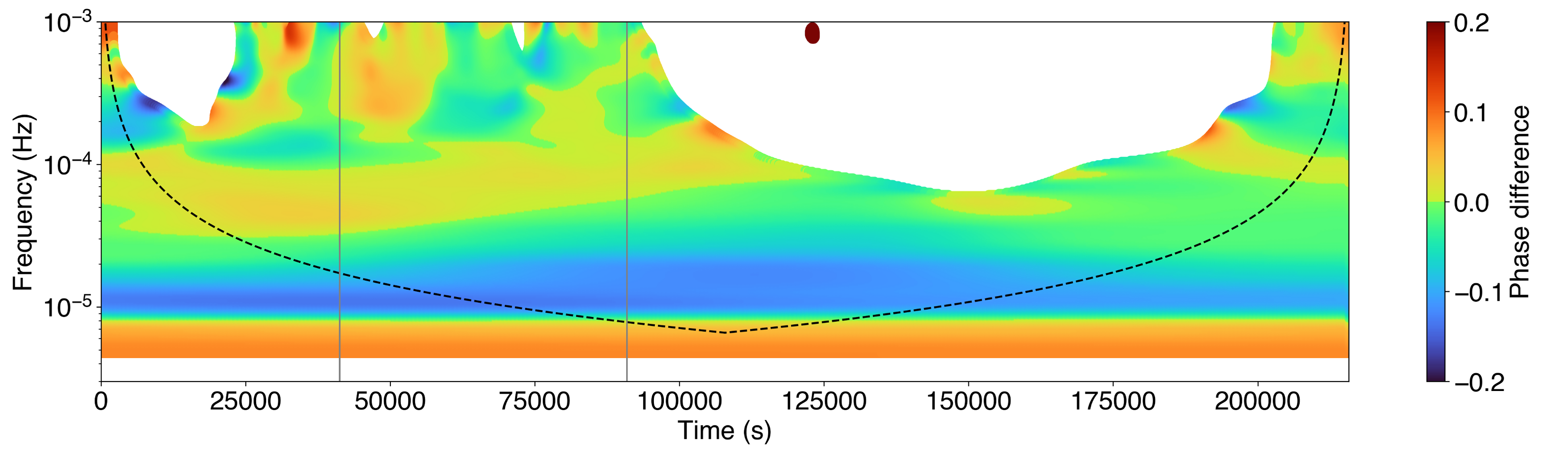}
    }
    \centerline{
        \includegraphics[height=0.25\textwidth]{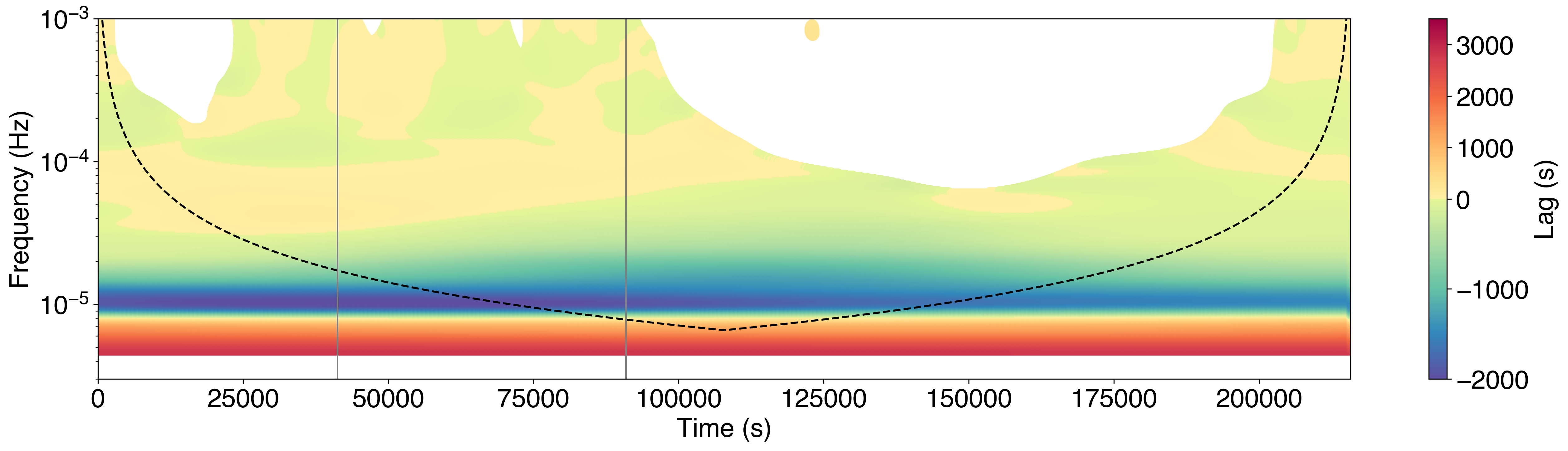}
    }
    \caption{Time-resolved phase differences (upper panel) and time lag (lower panel) between the soft (0.3--1 keV) and hard (1--4 keV) X-ray light curves of Fairall~9, c3. We exclude regions where the signal coherence falls below threshold values of 0.9. Plot details follow those described in the caption of Fig.~\ref{fig-phase-dif-sim}.}
    \label{fig-phase-dif}
\end{figure*}

Fig.~\ref{fig-phase-dif} represents the relative phase shift and the resolved time lags between the two energy bands of Fairall~9, c3. Initially, we used a coherence threshold to examine the phase structure throughout the time-frequency domain. We set the threshold to coherence $\geq 0.9$ to focus on the most reliable and high coherent regimes.\footnote{A coherence threshold of 0.7 was adopted in the simulations (Section 5) to allow the phase differences to be clearly visible, which is also appropriate for some sources such as IRAS 13224-3809 \citep{Wilkins2023}. For Fairall 9, which exhibits generally higher coherence, the threshold was adjusted to 0.9 to reflect the higher correlation significance in the data.} Comparing to the wavelet-lag analysis of IRAS13224--3809 reported by \citet{Wilkins2023}, where negative phase differences indicative of soft lags appear at high frequencies around $10^{-3}$ Hz, Fairall9 exhibits such negative phase differences at lower frequencies of $\sim 10^{-5}$ Hz, consistent with longer intrinsic variability timescales expected for its more massive black hole. Notably, the phase differences in Fairall~9, c3 are quite uniform at frequencies $\sim 8 \times 10^{-6}$--$3 \times 10^{-5}$~Hz, but their amplitudes are lower than those found in IRAS~13224--3809, suggesting weaker or more diluted reverberation signals.

\begin{figure*}
    \centering
    \includegraphics[width=\textwidth]{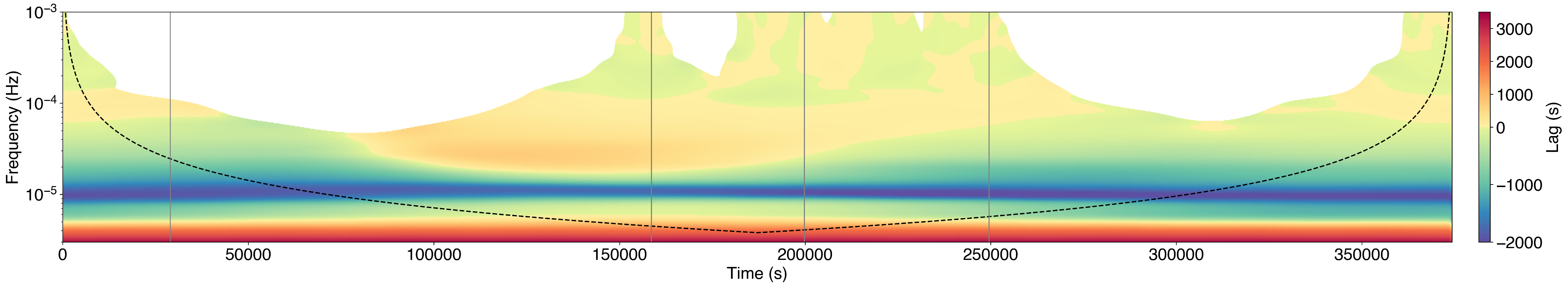}
    \caption{Same as Fig.~\ref{fig-phase-dif}, but for Fairall9, c5.}
    \label{fig-phase-dif-c5-f9}
\end{figure*}

In the case of Fairall~9, c5, the wavelet-based analysis reveals less-persistent lags at frequencies $\sim 5 \times 10^{-6}$--$5 \times 10^{-5}$~Hz (Fig.~\ref{fig-phase-dif-c5-f9}). Transient structures with varying amplitudes and temporal locations are evident, and localized regions where the soft-lag amplitude fluctuates appear at intermediate frequencies of $2$--$4 \times 10^{-5}$ Hz, particularly near the central part of the time series. These fluctuations coincide with the transition between the first three observations (Obs. ID 0605800401, 0101040201, and 0721110101), suggesting possible variability in the source geometry or physical conditions over time.

\begin{figure*}
    \centering
    \includegraphics[height=0.25\textwidth]{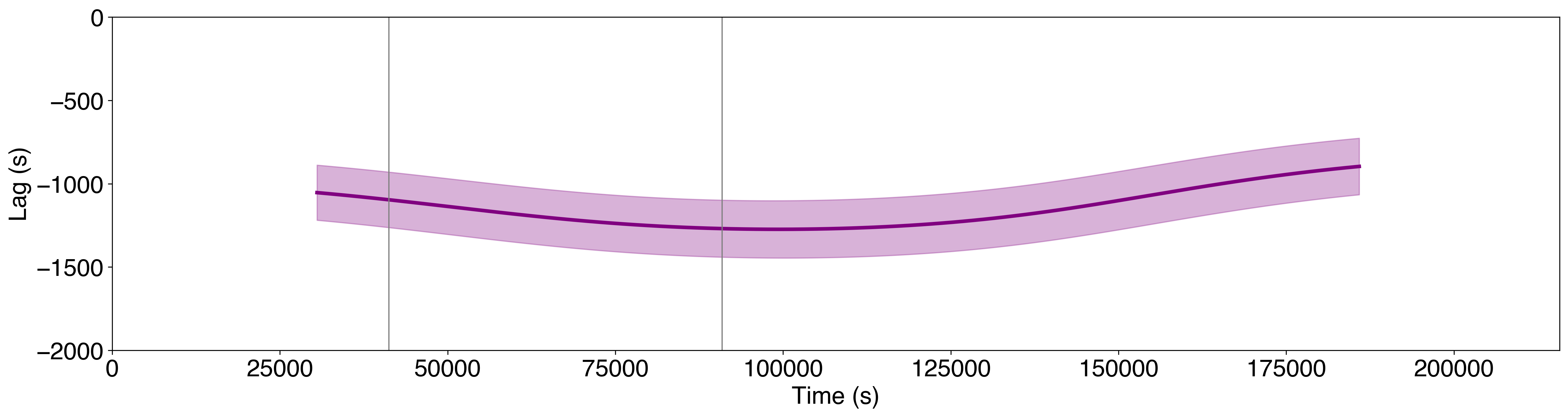}
    \caption{The averaged time lag between 0.3-1 keV and 1-4 keV bands for Fairall 9, c3. The lag profile is calculated and displayed only when the corresponding temporal and frequency bins intersect or lie within the COI.}
    \label{fig-avg-lag}
\end{figure*}

We now focus on Fairall 9, c3, whose light curve shows more persistent soft lags and was observed within $\sim 6$~months. Fig.~\ref{fig-avg-lag} presents the wavelet lag, averaged over $9\times10^{-6}$–$2.5\times10^{-5}$ Hz, as a function of time. The mean lag amplitude and its uncertainty, $\Delta\tau_{\rm lag}=\Delta\phi/(2\pi f)$, where $\Delta\phi$ is the phase-lag uncertainty (see, e.g., eq.~9 in \citealt{Wilkins2023}), are shown. At these frequencies, coherence remains high ($\gtrsim 0.9$) and quite stable. The lag curve and its associated uncertainty vary slightly over time, showing mean soft-lag amplitudes of $\sim 1000$~s. 

As most of the individual observations are too short to get into the low frequencies we want,  we compared the mean time lag from the final individual observation (Obs ID 0741330101) with that obtained from the stitched light curves Fairall~9, c3. In the single observation, lag amplitudes of $\sim 600$--800~s are detectable at $\sim 1$--3 $\times 10^{-5}$~Hz, still within the COI. Although larger lags of $\sim 900$--1000~s appear at slightly lower frequencies, they are outside the COI. In contrast, the stitched light curves Fairall~9, c3, extend the frequency coverage down to $\sim 9 \times 10^{-6}$~Hz, where the soft lags continue to increase from $\sim 600$~s up to $\sim 1000$~s, while remaining within the COI. The difference in lag amplitude between stitched and unstitched cases within overlapping frequency ranges is relatively small, which is $\sim 100$~s in our case. However, we note that the main difference is that stitching enhances access to lower-frequency, COI-valid regimes, where soft lags of $\sim 1000 \pm 200$~s are recovered.

\section{Discussion}

The interpretation of the soft excess as warm Comptonization is well established (e.g., \citealt{Noda2013, Petrucci2018}). While the spectra of Fairall~9 can be explained within various frameworks (e.g., \citealt{Liu2020}), we provide an additional analysis of its spectral fits using a hybrid corona model that combines both hot and warm corona, implemented by {\sc reXcor} model \citep{Xiang2022, Ballantyne2024}. The relativistic reflection features such as broadened Fe K line are notably weak or altogether absent \citep{Yaqoob2016}. One explanation for the absence of a broad line was that the black hole might have a truncated inner disc or low spin \citep[e.g.][]{Patrick2011, Hagen2023-Feb}. However, our spectral fitting results show that Fairall~9 can harbor a rapidly spinning black hole with an accretion disc extending to $r_{\rm ISCO}$. The spectral model also requires the hot lamp-post corona to be located at a moderate height, with a substantial fraction of the accretion energy dissipated in a warm, optically thick corona rather than solely in the hot corona. The requirement for the disc to extend down to $r_{\rm ISCO}$ aligns with some spectral models presented in previous literature \citep[e.g.,][]{Lohfink2012,Lohfink2016, Liu2020}. To test for ionized absorption in Fairall~9, we added a {\sc absori} warm absorber component in the spectral model. The fits show no significant improvement in fit statistics, suggesting the warm absorber is not required. The model further support the classification of Fairall~9 as a bare AGN \citep{Emmanoulopoulos2011, Patrick2011, Lohfink2016}.

Recent studies suggest that the warm corona model offers a more physically plausible explanation for the soft X-ray excess in AGN than reflection-only models, which often require extreme relativistic blurring or supersolar Fe abundances \citep[e.g.][]{Liu2020}. Based on an \emph{eROSITA} survey, \cite{Waddell2024} found that warm corona models were statistically preferred over blurred reflection in most AGN with strong soft excesses. While Fairall 9 fits naturally into this category, our spectral modeling supports the hybrid corona model, indicating that the soft band likely arises from a combination of reflection and warm Comptonization. The fitted warm corona parameters are in line with those found in other Seyferts \citep{Petrucci2018, Ballantyne2024}. This suggests a potentially universal structure: an inner sandwich warm-corona on the disc that generates the soft excess, alongside a more centrally concentrated hot corona that produces the hard X-rays. However, the location of the hot lamp-post corona in Fairall~9 cannot be well constrained, either models with $h=5~r_{\rm g}$ or $h=20~r_{\rm g}$ can reasonably fit the data. \citet{Liu2020} analyzed Fairall~9 spectrum using {\it XMM-Newton} and {\it NuSTAR} data, finding that a {\sc relxill}-only model poorly fit the data. A better fit was obtained by combining {\sc nthcomp} with {\sc relxill}. However, their study highlights that there is probably no unique interpretation using only spectral data.

Traditional Fourier lag analysis has struggled to identify significant lags in high-mass AGN due to their longer intrinsic reverberation timescales. These reverberation timescales scale approximately linearly with black hole mass \citep[e.g.][]{DeMarco2013, Kara2016, Mallick2021} and are thus pushed to lower Fourier frequencies for massive systems like Fairall~9. Furthermore, Fourier-based lag analysis might fail to detect the robust reverberation lags if the light curves are short and the lag is not coherent across the full light curve. The reverberation signal can be smeared by non-stationary lag evolution (e.g. if the geometry of the corona and disc system varies over time, producing reverberation signals that come and go) or buried in noise. This makes Fourier-based lag analysis particularly challenging for sources like Fairall~9, where the observations are short and the lag signal may be inherently weak or variable.

Meanwhile, wavelet methods being more sensitive to short-duration lags than standard Fourier analysis, which averages out transient signals. In Fairall~9, c3, we find high coherence of $\gtrsim 0.9$ over several time intervals and frequencies, particularly between $\sim 8 \times 10^{-6}$—$6 \times 10^{-5}$~Hz, where the phase differences evolve dynamically. The corresponding time lags range from hard (positive) lags at the lowest frequencies to significant soft (negative) lags of $\sim 1000$~s at frequencies of $\sim 9\times 10^{-6}$--$2.5 \times 10^{-5}$~Hz. At higher frequencies, beyond $\sim 10^{-4}$~Hz, we observe that the lags become more ambiguous, with fluctuations around zero and reduced phase coherence. This behavior is suggestive of phase wrapping, a known effect in high-frequency lag measurements where the periodic nature of the Fourier phase introduces artificial oscillations \citep{Nowak1999, Uttley2014}. These frequency-dependent positive and negative lags likely reflect a combination of inwardly propagating accretion rate fluctuations \citep[e.g.][]{Arevalo2006} and reverberation lags caused by the reprocessing of primary X-rays in the inner disc \citep{Uttley2014, Cackett2021}.


The comparative wavelet analysis with IRAS~13224–3809 that contains a much smaller black hole mass of $\sim 2 \times 10^{6} M_{\odot}$ \citep{Alston2020} helps contextualize our results within the broader AGN mass-lag scaling framework. In IRAS~13224–3809, X-ray soft lags are observed at higher frequencies of $\sim 10^{-3}$~Hz, which is consistent with expectations given its smaller gravitational timescales and is in agreement with previous studies using traditional lag-frequency analysis \citep[e.g.][]{Kara2013, Alston2020}. Interestingly, the coherence of signals on the reverberation timescales of IRAS~13224–3809 are quite low \citep{Wilkins2023}, compared to the case of Fairall~9. The observed variability in wavelet coherence is probably one of the reasons why the reverberation lag is not always detectable. Based on the results presented in Fig.~\ref{fig-wavelet-coh} and Fig.~\ref{fig-phase-dif}, we observe a highly correlated negative lag in Fairall~9. The negative lags especially observed in Fairall~9, c3 appear considerably more stable over time compared to those in IRAS~13224--3809, where the reverberation lags exhibit rapid, complex variability and relatively low coherence. This suggests a more stable source geometry in Fairall~9.

Nevertheless, for Fairall~9, c5, transient soft X-ray lags can still be found primarily in the frequency range of $\sim 5 \times 10^{-6}$--$5 \times 10^{-5}$~Hz. These soft lags are varied between 500--1500~s, indicating a more complex or evolving geometry during some particular periods. Such evolving geometry has been reported in other AGN through different methods. For example, spectral fitting and analysis of the emissivity profile has shown that jet-like structures can collapse into a compact X-ray corona \citep{Wilkins2015}. Additionally, Granger-causality tests have revealed potential multiple or time-varying lag components, indicating that AGN coronal heights may evolve even during a single observation \citep{Chainakun2023, Nakhonthong2024}. While \citet{Lohfink2016} proposed that Fairall~9 maintains a relatively stable coronal geometry, our findings suggest that such stability appears limited to specific epochs, particularly the last three observations with the strongest soft excess (Fairall~9, c3), where the lag signatures persist more consistently.

In principle, short-timescale wavelet coherence can offer valuable insight into AGN variability. Previous studies show that coherence between hard and soft X-ray bands stays high at low frequencies but drops sharply beyond a break frequency \citep[e.g.,][]{Epitropakis2017}. Reverberation models link this to phase-wrapping and geometry-dependent coronal variability \citep{Emmanoulopoulos2014,  Wilkins2016,Cackett2021}. Hence, variations in the highest frequency at which coherence remains significant may indicate changes in the intrinsic lag amplitude, driven by coronal fluctuations or inner accretion flow variations that shift the phase-wrapping frequency and the resulting coherence cut-off. Notably, only observations O3 and O4 retain robust coherence near $\sim 5 \times 10^{-4}$~Hz (i.e., timescales of $\sim 2000$~s), suggesting that during these epochs the emitting regions may have been more compact or better coupled, allowing coherent variability to persist to shorter timescales. However, reduced coherence can also result from additional factors such as dilution by the direct continuum, variations in scattering or reprocessing geometry, and stochasticity in the accretion flow, all of which can weaken phase coupling across energy bands. These effects, combined with the limited duration of the available light curves, mean that robust constraints on short-timescale behavior remain elusive, particularly for a high-mass AGN like Fairall~9 where reverberation signals are inherently shifted to lower frequencies.

\begin{figure}
        \centering
        \begin{subfigure}[h]{0.50\textwidth}
            \centering
            \includegraphics[width = 1. \textwidth]{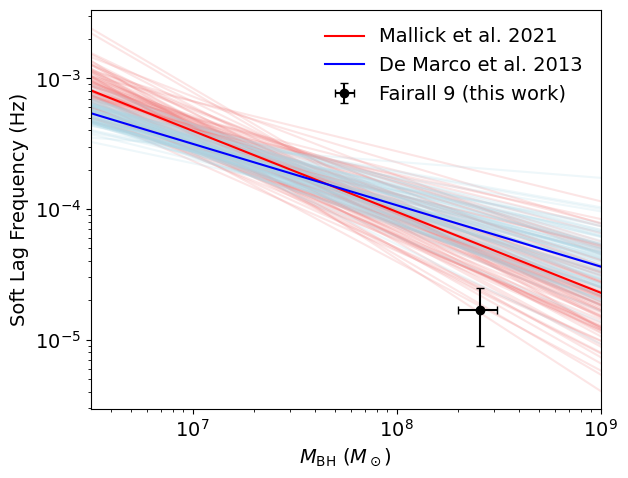}
        \end{subfigure}
        ~
        \begin{subfigure}[h]{0.50 \textwidth}
            \centering
            \includegraphics[width = 1.\textwidth]{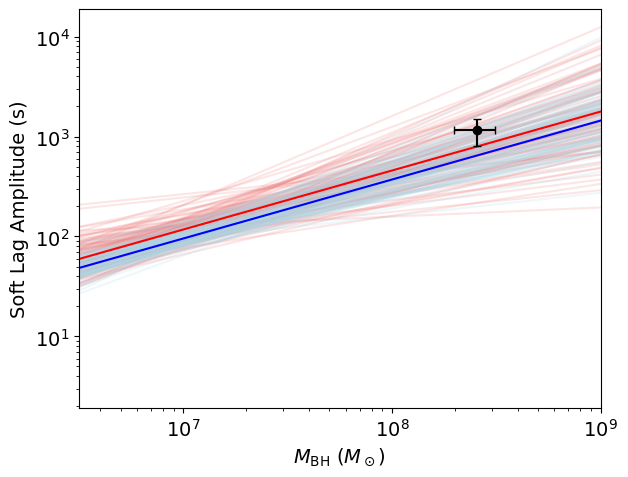}
        \end{subfigure}
        \caption{Soft lag frequency versus black hole mass (upper panel) and soft lag amplitude versus black hole mass (lower panel), as derived within the observational X-ray reverberation framework. The blue solid lines represent the best-fit linear regression models derived from 15 AGN in \citet{DeMarco2013}, with the light blue shaded regions indicating the corresponding confidence intervals based on the reported errors. The red lines and shaded regions show the relations obtained from 21 AGN sources in \citet{Mallick2021} where low mass AGN were also included. The black data point denotes the soft lag frequency and amplitude measured for Fairall~9 in this work.}
        \label{fig-lag-compare}
    \end{figure}

Our analysis of Fairall 9, c3, provides evidence for X-ray reverberation, though still not as definitive as seen in lower-mass AGN. Fig.\ref{fig-lag-compare} presents the relationship between soft X-ray lag properties and black hole mass, as previously reported by \citet{DeMarco2013} and \citet{Mallick2021}. Overlaid on these empirical relations are the lag properties of Fairall~9, as estimated in this work. Fairall~9, c3, lies at the high-mass end of the distribution and exhibits a relatively low-frequency yet high-amplitude lag. Nevertheless, the hint of soft lags in Fairall~9 fall within the expected range defined by the broader AGN sample, showing plausible agreement with the established trends. This consistency supports the interpretation that the soft lags in Fairall~9 possibly originate from X-ray reverberation processes, governed by light-travel delays between the compact hot corona and the inner regions of the accretion disc. Its geometry and physical scale are broadly aligned with expectations from standard reverberation models, where black hole mass scaling plays an important role.

While the presence of a warm corona can modify the soft-band response, the wavelet-detected soft X-ray lags align well with the {\sc reXcor} spectral model. The hot and warm corona act as a coupled, dynamic system that affects light-travel times and shapes the observed X-ray reverberation. We estimate that relativistic reflection plus warm-Comptonization flux contributes up to $\sim 0.5$ of the continuum below 1 keV, which is still strong enough to produce detectable reverberation signals and maintain high coherence. The $\sim 1000$~s negative lags likely occur as hot-corona variations drive delayed responses from the warm corona and inner disc. A closer examination of the spectral fitting results (Table~\ref{tab:bestfit_a099_l01}) reveals subtle yet notable differences between the first two and the last three observations of Fairall~9. Early epochs exhibit large uncertainties and stronger variability in key warm-corona parameters such as the heating fraction ($h_{\rm f}$) and optical depth ($\tau$), regardless of coronal height. In contrast, the final three observations display more consistent and well-constrained values, with $\Gamma$ tightly clustered around 1.90--2.05, and $\tau$ confined to narrower ranges of $\sim 14$--24 (for $h = 5~r_{\rm g}$) and 10--14 (for $h = 20~r_{\rm g}$). These trends of more stable spectral parameters coincide with less variable soft-band reverberation, and, consequently, imply a more stable geometry during the last three observations.

One crucial test for the existence of the warm corona could be revealed by observations from the Imaging X-ray Polarimetry Explorer (\emph{IXPE}; \citealt{Odell2018}). Given the proposed additional warm, optically thick corona geometry, a deviation in polarization degree and angle is expected when compared to the traditional single-corona geometry \citep{Ursini2022,Marinucci2019,Tagliacozzo2023}. Polarimetric data should therefore provide valuable information to constrain the AGN geometry responsible for producing the X-ray emission.

\section{conclusion}

We perform spectral and wavelet analysis of Fairall~9. The results show that the hot coronal height is not well constrained, i.e., the fits are reasonably good with either $h=5~r_{\rm g}$ or $20~r_{\rm g}$, which are the discrete options available in {\sc reXcor}. The soft excess is primarily attributed to warm Comptonization while distant reflection also contributes significantly. The spectral modeling supports a scenario in which both hot and warm corona coexist, assuming a high black hole spin of $a = 0.99$. Under this condition, the disc can extend close to the event horizon, enabling relativistically blurred inner-disc reflection. We find evidence for transient soft X-ray lags that are localized in both time and frequency domains. More persistent soft lags are observed in the Fairall~9, c3 dataset. Given the mass and structure of Fairall~9, the lags are plausibly consistent in amplitude ($\sim 1000$~s) and timescales ($\sim 9\times 10^{-6}$--$2.3 \times 10^{-5}$~Hz) with expectations from X-ray reverberation. Future work should extend this approach to larger samples of high-mass AGN and explore the effects of coronal geometry variability and warm corona evolution. Extended monitoring with \emph{XMM-Newton} or upcoming observatories like \emph{NewAthena}, particularly in coordination with \emph{IXPE}, will enable more definitive mapping and modeling of disc-corona coupling in massive AGN such as Fairall~9.

\section*{Acknowledgements}

We thank the referee for the helpful comments that have improved the quality of our paper. This project is funded by National Research Council of Thailand (NRCT) and Suranaree University of Technology, grant number N42A680156. PC also thanks (i) Suranaree University of Technology (SUT), (ii) Thailand Science Research and Innovation (TSRI), and (iii) National Science, Research and Innovation Fund (grant number 204265). KK acknowledges the support from the Institute for the Promotion of Teaching Science and Technology (IPST) of Thailand. We confirm that there are no conflicts of interest associated with this submission.

\section*{Data availability}

The data analyzed in this article are available from the corresponding author upon reasonable request. The {\sc reXcor} spectral models employed in this analysis are publicly accessible through the official {\sc xspec} website.


\bibliographystyle{mnras}

\bsp	
\label{lastpage}
\end{document}